\documentclass{pasa}%

\usepackage{hyperref}
\hypersetup{
    colorlinks=true,
    linkcolor=blue,
    filecolor=magenta,      
    urlcolor=cyan,
}

\usepackage{graphicx}	
\usepackage{amsmath}	
\usepackage{amssymb}	
\usepackage{multicol}        
\usepackage{bm}		
\usepackage{pdflscape}	
\usepackage{threeparttable}
\usepackage{txfonts}

\usepackage{layouts}

\usepackage[dvipsnames]{xcolor}

\title[Fornax cluster magnetic fields]{Early Science from POSSUM: Shocks, turbulence, and a massive new reservoir of ionised gas in the Fornax cluster}

\author[C.~S. Anderson et al.]{C.~S. Anderson$^{1,2,3}$,
G.~H. Heald$^{2}$,
J.~A. Eilek$^{1,4}$,
E. Lenc$^{3}$,
B.~M. Gaensler$^{5}$,
Lawrence Rudnick$^{6}$,
C.~L. Van Eck$^{5}$,
S.~P. O'Sullivan$^{7}$,
J.~M. Stil$^{8}$,
A. Chippendale$^{3}$,
C.~J. Riseley$^{9,10}$,
E. Carretti$^{10}$,
J. West$^{5}$,
J. Farnes$^{11}$,
L. Harvey-Smith$^{12,13}$,
N.~M. McClure-Griffiths$^{14}$,
Douglas C.~J. Bock$^{3}$,
J.~D. Bunton$^{3}$,
B. Koribalski$^{3,13}$,
C.~D. Tremblay$^{2}$,
M.~A. Voronkov$^{3}$,
K. Warhurst$^{2}$
\affil{$^{1}$Jansky fellow of the National Radio Astronomy Observatory, 1003 Lopezville Rd, Socorro, NM 87801 USA}
\affil{$^{2}$CSIRO Astronomy and Space Science, PO Box 1130, Bentley WA 6102, Australia}
\affil{$^{3}$ATNF, CSIRO Astronomy and Space Science, PO Box 76, Epping, New South Wales 1710, Australia}
\affil{$^{4}$Physics Department, New Mexico Tech, Socorro NM 87801 USA}
\affil{$^{5}$Dunlap Institute for Astronomy and Astrophysics, University of Toronto, 50 St. George Street, Toronto, ON M5S 3H4, Canada}
\affil{$^{6}$Minnesota Institute for Astrophysics, University of Minnesota, 116 Church St. SE, Minneapolis, MN 55455 USA}
\affil{$^{7}$School of Physical Sciences and center for Astrophysics \& Relativity, Dublin City University, Glasnevin, D09 W6Y4, Ireland}
\affil{$^{8}$Department of Physics \& Astronomy, The University of Calgary, 2500 University Drive NW, Calgary AB, T2N 1N4, Canada}
\affil{$^{9}$Dipartimento di Fisica e Astronomia, Universit\`{a} degli Studi di Bologna, via P. Gobetti 93/2, 40129 Bologna, Italy}
\affil{$^{10}$INAF - Istituto di Radioastronomia, Via Gobetti 101, 40129 Bologna, Italy}
\affil{$^{11}$Oxford e-Research center (OeRC), Department of Engineering Science, University of Oxford, Oxford, OX1 3QG, UK}
\affil{$^{12}$School of Physics, University of New South Wales, Sydney, NSW 2052, Australia}
\affil{$^{13}$Western Sydney University, Locked Bag 1797, Penrith, NSW, 2751, Australia}
\affil{$^{14}$Research School of Astronomy \& Astrophysics, Australian National University, Canberra ACT 2611 Australia}}

\jid{PASA}
\doi{10.1017/pas.\the\year.xxx}
\jyear{\the\year}

\usepackage{aas_macros}
\usepackage{hyperref} 
\hypersetup{colorlinks,citecolor=blue,linkcolor=blue,urlcolor=blue}

\begin{document}

\begin{frontmatter}
\maketitle

\begin{abstract}
We present the first Faraday rotation measure (RM) grid study of an individual low-mass cluster --- the Fornax cluster --- which is presently undergoing a series of mergers. Exploiting commissioning data for the POlarisation Sky Survey of the Universe's Magnetism (POSSUM) covering a $\sim34$ square degree sky area using the Australian Square Kilometre Array Pathfinder (ASKAP), we achieve an RM grid density of $\sim25$ RMs per square degree from a 280 MHz band centred at 887 MHz, which is similar to expectations for forthcoming GHz-frequency $\sim3\pi$-steradian sky surveys. These data allow us to probe the extended magnetoionic structure of the cluster and its surroundings in unprecedented detail. We find that the scatter in the Faraday RM of confirmed background sources is increased by $16.8\pm2.4$ rad m$^{-2}$ within 1 degree (360 kpc) projected distance to the cluster centre, which is 2--4 times larger than the spatial extent of the presently-detectable X-ray-emitting intracluster medium (ICM). The mass of the Faraday-active plasma is larger than that of the X-ray-emitting ICM, and exists in a density regime that broadly matches expectations for moderately-dense components of the Warm-Hot Intergalactic Medium. We argue that forthcoming RM grids from both targeted and survey observations may be a singular probe of cosmic plasma in this regime. The morphology of the global Faraday depth enhancement is not uniform and isotropic, but rather exhibits the classic morphology of an astrophysical bow shock on the southwest side of the main Fornax cluster, and an extended, swept-back wake on the northeastern side. Our favoured explanation for these phenomena is an ongoing merger between the main cluster and a sub-cluster to the southwest. The shock's Mach angle and stand-off distance lead to a self-consistent transonic merger speed with Mach 1.06. The region hosting the Faraday depth enhancement also appears to show a decrement in both total and polarised radio emission compared to the broader field. We evaluate cosmic variance and free-free absorption by a pervasive cold dense gas surrounding NGC 1399 as possible causes, but find both explanations unsatisfactory, warranting further observations. Generally, our study illustrates the scientific returns that can be expected from all-sky grids of discrete sources generated by forthcoming all-sky radio surveys.
\end{abstract}

\end{frontmatter}



\section{Introduction}\label{sec:intro}

The Universe's baryons are mostly located outside the stellar envelopes of galaxies, in the vast expanses occupied by clusters of galaxies, and in filaments of tenuous plasma that connect them. The properties of gas in these different regimes --- including their state of magnetisation --- lie at the heart of theories of cosmic evolution and ecology, but remain difficult to pin down observationally. 
For instance, the hot intracluster medium (ICM) contains about 4\% of the baryonic mass of the late-time Universe, which is (for example) a higher proportion than is contained in stars \citep{deGraaff2019}. Magnetic fields break the isotropy of viscosity, pressure support, and thermal conductivity of the ICM, thereby exerting an out-sized influence on cluster physics and evolution. They can trace ordered and turbulent flows in plasma (e.g. \citealp{Anderson2018}), reveal interactions between the ICM and in-falling gas (e.g. \citealp{Keshet2017}), embedded galaxies (e.g. \citealp{DP2008,PD2010}) and galactic outflows (e.g. \citealp{Guidetti2011,Guidetti2012,Anderson2018}), and help reveal how the broader cosmos became magnetised (e.g. \citealp{Vazza2014,Bonafede2015}). 

Beyond the ICM, the Warm-Hot Intergalactic Medium (WHIM) must contain around 80\% of the Universe's baryons \citep{deGraaff2019}, though this material is comparatively unstudied, with only recent claims of detection of its sparser phases \citep{Nicastro2017,deGraaff2019,Tanimura2019,Macquart2020}. Simulations suggest that the WHIM will be found in a diverse set of regimes, occupying relatively dense agglomerations ($\delta\sim200$, where $\delta$ is the over-density factor of baryons compared to the cosmic mean, defined by $\delta\equiv\rho/\bar{\rho}-1$, and in turn, $\rho$ is the baryon number density at a given location, while $\bar{\rho}$ is the mean baryon number density in the Universe, which is currently $\bar{\rho}\approx2\times10^{-7}$ cm$^{-3}$; \citealp{Planck2016}) around galaxy clusters in its densest and hottest manifestations, and in tenuous filaments between massive galaxies in its sparsest and coolest regimes ($\delta\sim$ a few) \citep{Dave2001}. The magnetisation of this material is also consequential, since the predictions of models for cosmic magneto-genesis differ most strongly here (e.g. \citealp{Donnert2018} and references therein), which is just now beginning to be revealed with extraordinary new low frequency radio measurements \citep{Govoni2019,Botteon2020}, and may also be probed via measurements of radio polarisation and Faraday rotation (e.g. \citealp{Akahori2018,Locatelli2018,OSullivan2020}). 

Faraday rotation is an effective tracer of the distribution and properties of rarefied, magnetised cosmic plasma, such as the ICM and WHIM (e.g. \citealp{CP1962,Burn1966,Conway1974, KS1976,TP1993,Farnsworth2011,OSullivan2013b,JH2015,Gaensler2015,Anderson2016b}). Consider that the linear polarisation state of radio emission can be described by a complex vector $\boldsymbol{P}$, related to the Stokes parameters $Q$ and $U$, the polarisation angle $\psi$, the fractional polarisation $p$ and the total intensity $I$ as

\begin{equation}
\boldsymbol{P} = Q + iU = pIe^{2i\psi}
\label{eq:ComplexPolVec}
\end{equation}

\noindent After being emitted at a location $L$, linearly polarised radiation will be Faraday rotated by magnetised thermal plasma along the line of sight (LOS) to an observer by an amount equal to

 \begin{equation}
\Delta\psi= \phi(L)\lambda^2
\label{eq:rotation}
\end{equation}

\noindent where $\lambda$ is the observing wavelength, and $\phi$ is the Faraday depth, given by  

\begin{equation}
\text{$\phi$}(L) = 0.812 \int_{L}^{0} n_e\boldsymbol{B}\cdot\text{d}\boldsymbol{s}~\text{rad m}^{-2}
\label{eq:FaradayDepth}
\end{equation}
 
\noindent and, in turn, $n_e$ [cm$^{-3}$] \& $\boldsymbol{B}$ [$\mu$G] are the thermal electron density and magnetic field along the LOS respectively. 

The observable polarisation spectrum $\boldsymbol{P}(\lambda^2)$ is obtained by summing the polarised emission emerging from all possible Faraday depths within the synthesised beam of the telescope:

 \begin{equation}
\boldsymbol{P}(\lambda^2) = \int_{-\infty}^{\infty} \boldsymbol{F}(\phi) e^{2i\phi\lambda^2} d\phi
\label{eq:SumPol}
 \end{equation}
 
The function $\boldsymbol{F}(\phi)$ (the so-called Faraday Dispersion Spectrum, henceforth FDS) specifies the distribution of polarised emission as a function of Faraday depth along the LOS. This quasi-Fourier relationship can be inverted to reconstruct $\boldsymbol{F}(\phi)$ given observations of $\boldsymbol{P}(\lambda^2)$ \citep{Burn1966,BdB2005}. In the common situation where a point-like source is viewed through an extended reservoir of magnetised thermal plasma lying in the foreground (which we deal with in this work), $\boldsymbol{F}(\phi)$ shows a single peak at a well-defined Faraday depth $\phi_\mathrm{peak}=\text{argmax}(|\boldsymbol{F}(\phi)|)$. This is then equivalent to the so-called Faraday rotation measure (RM) of the source. Largely for historic reasons, we continue to use the term ``RM'' in places in this work, though the measurements themselves are of $\phi_\mathrm{peak}$ --- i.e. extracted through the method of RM synthesis, rather than from the gradient of the polarisation angle as a function of $\lambda^2$. Regardless of nomenclature, $\phi_\mathrm{peak}$ provides a direct measure of the amount of magnetised thermal plasma that has been traversed along the line of sight. Ensembles of polarised radio sources can therefore back-illuminate the magnetoionic structure of extended plasma reservoirs in the foreground. 

%


Applying these `RM grid' techniques to study the ICM or WHIM in individual galaxy clusters has historically been difficult, because past generations of radio instrumentation could only recover a relatively low density of polarised sources over the required multi-square-degree sky areas ($\mathcal{O}(1)$ linearly polarised source per square degree --- e.g. \citealp{TSS2009}), combined with the uncertainty of whether any given RM grid source is located behind a target cluster, inside it, or in the foreground. Fortunately, these limitations will soon be transcended using data from a new generation of radio interferometers and optical redshift surveys. In the former domain, the Australian Square Kilometre Array Pathfinder (ASKAP) can routinely measure polarised source densities of $\sim25$ per square degree over tens of square degrees in just a few hours (this work, West et al. \emph{in prep.}), and will soon survey the entire southern sky at this or greater depth for the Polarisation Sky Survey of the Universe's Magnetism (POSSUM; \citealp{Gaensler2010}). At the same time, deep pointed observations like the MeerKAT \citep{Jonas2009} Fornax Survey \citep{Serra2016} may soon recover $\sim$hundreds of polarised sources per square degree. Both approaches will open up the field of observational galaxy cluster astrophysics to routine RM grid studies \citep{Heald2020}. 

This paper heralds the dawn of this era by exploiting ASKAP's unique imaging capabilities to study the ionised gas in the Fornax cluster. The Fornax cluster is nearby (20.64 megaparsecs; Mpc; \citealp{Lavaux2011}), but much poorer than similarly well-studied clusters like the Virgo and Coma clusters. It has only $\sim390$ member galaxies (which are typically low in mass; \citealp{Maddox2019}), and has a comparatively low total mass of $6^{+3}_{-1}\times10^{13}$ M$_\odot$ \citep{Drinkwater2001,Nasonova2011,Maddox2019}, which is one and two orders-of-magnitude less massive than the Virgo and Coma clusters respectively. The (presently detectable) X-ray-emitting ICM is also small, extending asymmetrically outward from NGC1399 to a mere 15\%--30\% of the cluster's 1.96 degree virial radius. Nevertheless, its halo mass is more representative of that in which the majority of the galaxies in the Universe reside and evolve (\citealp{Haan2014} and references therein). The dynamical state of the cluster, and associated astrophysical processes, are therefore of keen interest.  

Our understanding of the dynamics of this system is evolving rapidly. The core of the Fornax cluster is densely populated with early-type galaxies that possess relatively uniform properties and low velocity dispersion. This used to be cited as evidence that the Fornax cluster is virialised and well-evolved (see \citealp{Iodice2019} and references therein), but it is now clear that the Fornax cluster possesses complex spatial sub-structure both in terms of its constituent gas and galaxies (e.g. \citealp{Drinkwater2001,Paolillo2002,Scharf2005,Su2017,Venhola2018,Sheardown2018}), and is still assembling mass through a series of on-going mergers. At the largest scales, \citet{Drinkwater2001} argue that there is a genuine mass partition between northeast and southwest sub-components of the cluster, which are respectively dominated by the cD-type galaxy NGC 1399 and NGC 1316 --- the host galaxy of the radio source Fornax A --- a few degrees ($\sim1.3$ Mpc) away. \citet{Scharf2005} cite the swept-back (to the northeast) morphology of the hot ICM traced by X-rays as possible evidence that these sub-components are merging at transonic speeds. On smaller scales, NGC 1399 is undergoing a series of close encounters with the spiral galaxy NGC 1404, as the latter falls into the cluster potential and interacts with the diffuse gas there \citep{Machacek2005}. This appears to have induced sloshing in the ICM which is most apparent in the central $\sim30$ kpc of the cluster \citep{Su2017}, and may have generated shocks and turbulence on scales up to more than a degree \citep{Sheardown2018}. 

Thus, the Fornax cluster differs greatly in its properties from other relatively nearby massive clusters, including Virgo and Coma, and from the massive clusters from which our canonical understanding of the magnetised ICM were chiefly derived. While our understanding of the magneto-thermal plasma structures of even large clusters remain incomplete (e.g. \citealp{JH2015,Heald2020}), our ignorance is much more pronounced at the important low-mass end of the halo distribution. Sensitive new RM grid experiments can directly reveal the magnetised gas in such environments, providing sorely needed new data in this sphere. In this work then, our overarching aim is to search for Faraday RM enhancements to trace the structure of magnetised ionised gas in an individual low-mass galaxy cluster for the first time. Our specific goals are to (a) estimate the mass and extent of any such material, and to compare these estimates to those determined for ensembles of the more massive clusters previously probed in RM stacking experiments, (b) to determine whether any material thereby revealed differs in its properties or extent from that revealed by Bremsstrahlung radiation, thus establishing the complementarity of the two measurement techniques, and (c) demonstrate what unique information the RM measurements provide about the dynamical mass assembly processes that continue to take place in the system. The spatial density and areal sky coverage of our RM grid, coupled the fact that it is dominated by confirmed background radio sources (as we will show), is groundbreaking in this field. 

Our paper is organised as follows. We describe our observations, calibration, and imaging procedures in Section 2, and our polarimetric analysis in Section 3. Analysis of ancillary redshift data is presented in Section 4. We present our results, discussion, and conclusions in Sections 5, 6 and 7 (respectively). In this paper, we adopt a distance to NGC 1399 of D = 20.64 Mpc \citep{Lavaux2011}, which yields an image scale of 100 parsecs/arcsec. Cardinal directions are referred to by their usual abbreviations --- e.g. NE, SW, S for northeast, southwest, and south, respectively. We use the spectral index convention $S\propto\nu^\alpha$.

\section{Observations, calibration and imaging}\label{sec:obs}

\begin{table*}
\caption{Summary of observations}
\begin{threeparttable}
\centering
\begin{tabular}{ll}
\hline
\hline
Target & Fornax cluster region\\
Scheduling block ID & 8279\\
Date of observations & 23rd March 2019 \\
Field centre (J2000) & $03^h29^m30^s$, $-34^d58^m30^s$ \\
Field centre (J2000, Gal. $l,b$) & 235.988$^\circ$, $-$55.484$^\circ$\\
No. telescope pointings & 1 \\
Total integration time & 6 hours \\
Full-band sensitivity $^\dagger$ & 30 $\muup$Jy beam$^{-1}$ \\
Recorded polarisations & $XX$, $XY$, $YX$, $YY$ \\
Beam footprint & {\tt square\_6x6} \\
Beam pitch & 0.9 deg\\
Number of valid beams & 30 paired X and Y pol. (of 36)\\
FWHM of formed beams at 887 MHz$^\ddagger$ & 1.59 deg\\
Total sky coverage & $\sim34$ deg$^2$ \\
Number of antennas & 32 \\
Antenna diameter & 12 m \\
Shortest baseline & 22.4 m \\
Longest baseline & 6.4 km \\
Angular resolution (robust $=0$) & 11$\times$14 arcsec \\
Largest recoverable angular scale$^{\ddagger\ddagger}$ & 30 arcmin\\
Frequency range (central frequency) & 747--1027 MHz (887 MHz)\\
Frequency resolution & 1 MHz \\
$\lambda^2$ range & 0.085--0.161 m$^2$ \\
Resolution in $\phi$ space $\textsuperscript{a}$ & 46 rad m$^{-2}$ \\
Largest recoverable $\phi$-scale$^{\ddagger\ddagger}$ $\textsuperscript{a}$ & 37 rad m$^{-2}$ \\
Largest recoverable |$\phi$| $\textsuperscript{a}$ & 510 rad m$^{-2}$ \\
\hline
\end{tabular}
\begin{tablenotes}
\small
\item $^\dagger$ Measured per Stokes parameter in multi-frequency synthesis images generated with a Briggs' robust weighting value of 0.0. $^\ddagger$ At centre frequency of band. $^{\ddagger\ddagger}$ At greater than 50\% sensitivity. $\textsuperscript{a}$ Calculated from equations in Section 6 of \citet{BdB2005}.
\end{tablenotes}
\end{threeparttable}
\label{tab:obsdeets}
\end{table*}

\begin{figure}
\centering
\includegraphics[width=0.5\textwidth]{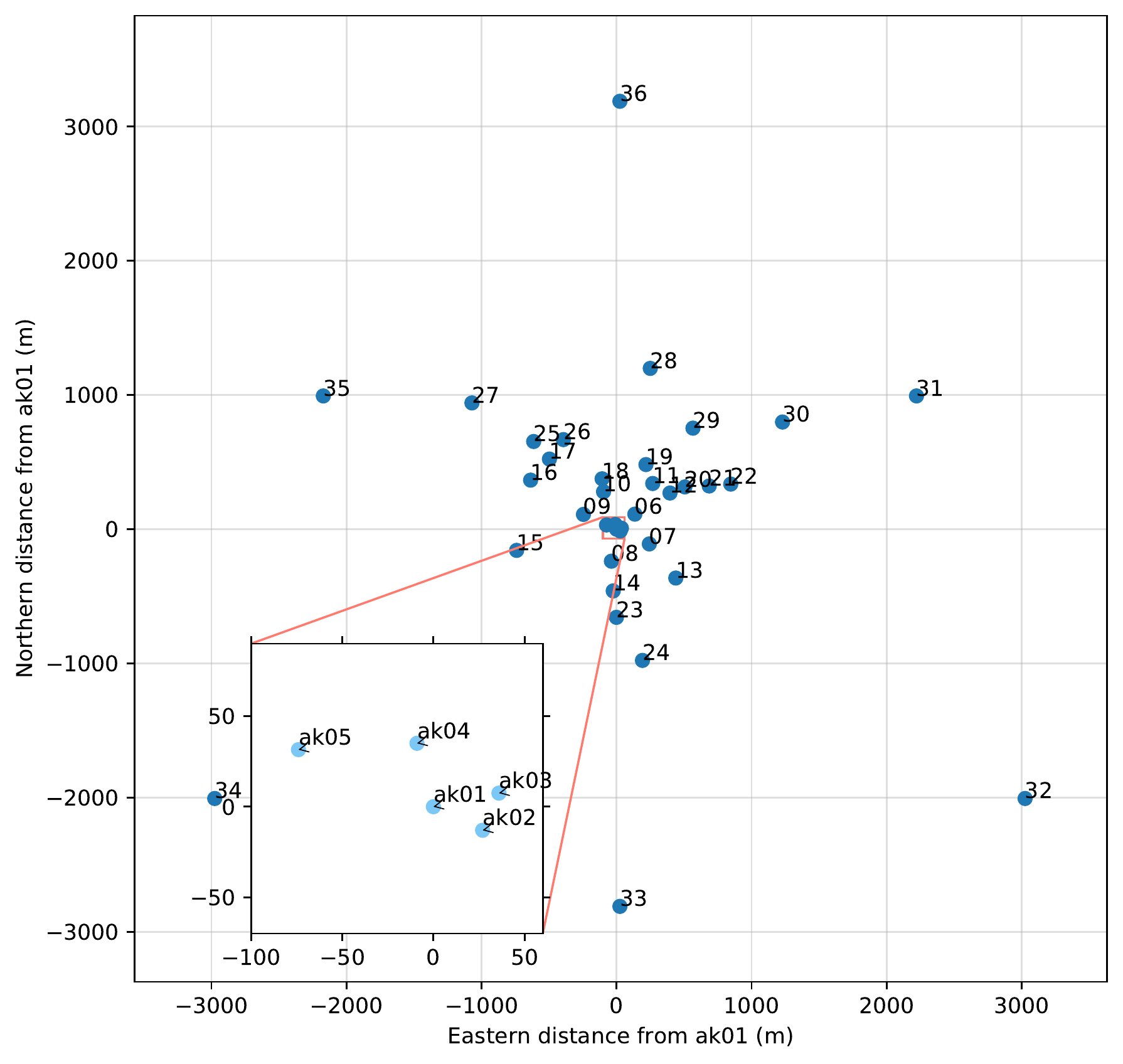}
\caption{Offset ASKAP antenna positions in meters from antenna ak01 (Longitude: 116.631424$^\circ$ E, Latitude: -26.697000$^\circ$; \citealp{McConnell2020}\protect\footnotemark) for the full ASKAP array. The inset panel zooms in on E-W offsets of -100m to +60m, and N-S offsets of -70m to +90m.}
\label{fig:array_config_fornax}
\end{figure}

\begin{figure}
\centering
\includegraphics[width=0.5\textwidth]{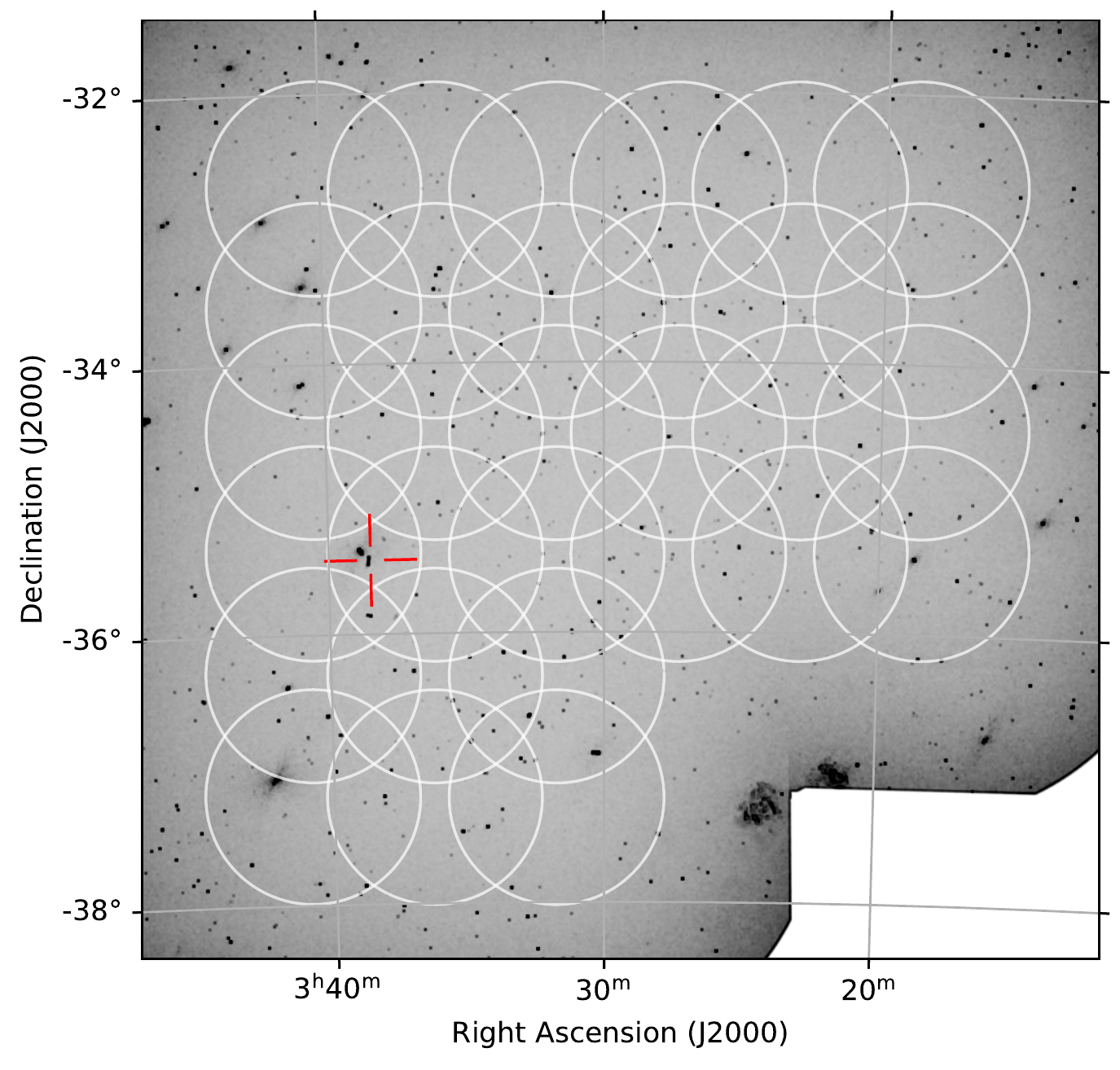}
\caption{The position (and FWHM at our maximum frequency of 1027 MHz) of formed ASKAP beams used in this work (white circles), overlaid on a map of the local root-mean-squared (RMS) noise in peak-$P$ (see Figure \ref{fig:peak_P_noise} and Section \ref{sec:pol_analysis} for an enlarged version containing further detail). The centre of the Fornax cluster is indicated with a red cross-hair, and the lobes of Fornax A are partially visible to the south west.}
\label{fig:peak_P_noise_beams}
\end{figure}

We observed the Fornax cluster region during commissioning tests of the ASKAP radio telescope \citep{DeBoer2009, Johnston2007, SB2016}. ASKAP consists of $36 \times 12$-m antennas, each  equipped with a Phased Array Feed (PAF), yielding a $\sim30$ square degree instantaneous field of view (depending on frequency) and high survey speed. We observed a single such pointing for six hours with a {\tt square\_6x6} beam footprint \citep{McConnell2016}, using a beam pitch (horizontal and vertical angular separation) of 0.9 degrees, covering the frequency range 747--1027 MHz (averaging to 1 MHz frequency channels on-the-fly from ASKAP's native 18 kHz spectral resolution), achieving a band-averaged sensitivity of 30 $\mu$Jy beam$^{-1}$ per Stokes parameter. The ASKAP array configuration is shown in Figure \ref{fig:array_config_fornax}. Of the 36 beams formed for our observations, six in the south-west corner of the mosaic suffered from beamforming errors leading to low sensitivity, and were discarded from further analysis. It is coincidental that the bright radio galaxy Fornax A appears in the vicinity of these beams for our observations. The sky position of the remaining valid beams are indicated in Figure \ref{fig:peak_P_noise_beams}. We note that the field centre was chosen to satisfy the competing demands of several science teams working in this sky area, and so while the field incorporates the Fornax Cluster, Fornax A and several other radio galaxies, it is not centred on any of them. Further details are summarised in Table \ref{tab:obsdeets}.

We flagged and calibrated our data in the Common Astronomy Software Applications ({\tt CASA}; \citealp{McMullin2007}) package. We  flagged radio frequency interference manually, which is feasible only because of the exceptionally RFI-quiet conditions at the Murchison Radio-astronomy Observatory (e.g. \citealp{Indermuehle2018}). We calibrated the flux scale, instrumental bandpass, and (on-axis) polarisation leakage `D-terms' using standard methods applied to observations of the (unpolarised) standard calibrator source PKS B1934--638. The frequency-dependent instrumental XY-phase was nulled out at the beamforming stage using the ASKAP on-dish calibration system (ODC; \citealp{Chippendale2019}). The off-axis polarimetric instrumental response was not corrected for this work. However, after the on-beam-axis D-term corrections are applied, and for the frequency and beam pitch employed, we estimate that the leakage from Stokes $I$ to $Q$ and $U$ is usually less than 1\%, though it can be worse in isolated areas. We discuss this more in Section \ref{sec:pol_analysis}. The absolute polarisation angle was also left uncalibrated, but the inter-beam and inter-channel relative polarisation angle is guaranteed to be consistent by the beamforming procedure, which we subsequently verified by comparing polarisation spectra of sources observed in images of adjacent beams. The absolute flux scale is uncertain by up to 10\%, and this can vary as a function of position in the field \citep{McConnell2020}.

\footnotetext{\url{https://confluence.csiro.au/download/attachments/733676544/ASKAP_sci_obs_guide.pdf}}

We imaged the data with {\tt WSClean} \citep{Offringa2014}. For all Stokes parameters, we generated image cubes with $5000\times5000$ pixels, a pixel scale of 2.5 arcseconds, a \citet{Briggs1995} robust weighting value of 0, local noise estimation with automatic CLEAN thresholding and masking (at 1 $\sigma$ and 3$\sigma$ respectively), and joined-channel CLEANing with 8 MHz channelisation. We performed two rounds of phase-only self-calibration using {\tt CASA} with a solution interval of 300 seconds, and then one round of phase and amplitude self-calibration with a solution interval of 60 seconds. We experimented with shorter solution intervals, but this produced little effect. We then re-imaged and {\sc clean}ed Stokes $I$ MFS maps independently, an then the Stokes $Q$ and $U$ datacubes using {\tt WSClean}'s `join polarisations' and `squared channel joining' modes. The individual beam images were then linearly mosaicked for all channels and Stokes parameters using the {\tt SWarp} package \citep{Bertin2002}, employing a scaled-width circular Gaussian beam, whose full-width-half-maximum scales as $(1.09/12)\lambda$ \citep{McConnell2020}\footnote{\url{https://confluence.csiro.au/download/attachments/733676544/ASKAP_sci_obs_guide.pdf}}, truncated at the 10\% power point. We then smoothed to the spatial resolution of our lowest frequency channel --- $18\times14$ arcseconds --- then re-gridded to a common pixel grid, and concatenated together to form Stokes I, Q and U datacubes with dimensions RA, Decl, $\lambda^2$.

We note that in the final linear mosaics, nine or more beams contribute to the data values at any point located inside the outer ring of beam centres, and that to our knowledge, the cluster centre is not in a `special' or otherwise noteworthy location with respect to the positions of the beam centres (see Figure \ref{fig:peak_P_noise_beams}).

\section{Polarimetric analysis}\label{sec:pol_analysis}

We calculated the Faraday Dispersion Spectrum (FDS) over the range -200 to +200 rad m$^{-2}$ using RM synthesis\footnote{\url{https://github.com/brentjens/rm-synthesis}, version 1.0-rc4} \citep{Burn1966,BdB2005} applied to the Stokes Q and U data cubes with equal weighting per image channel. The result is a complex-valued FDS datacube with dimensions RA, Decl, and $\phi$.  

\begin{figure*}
\centering
\includegraphics[width=\textwidth]{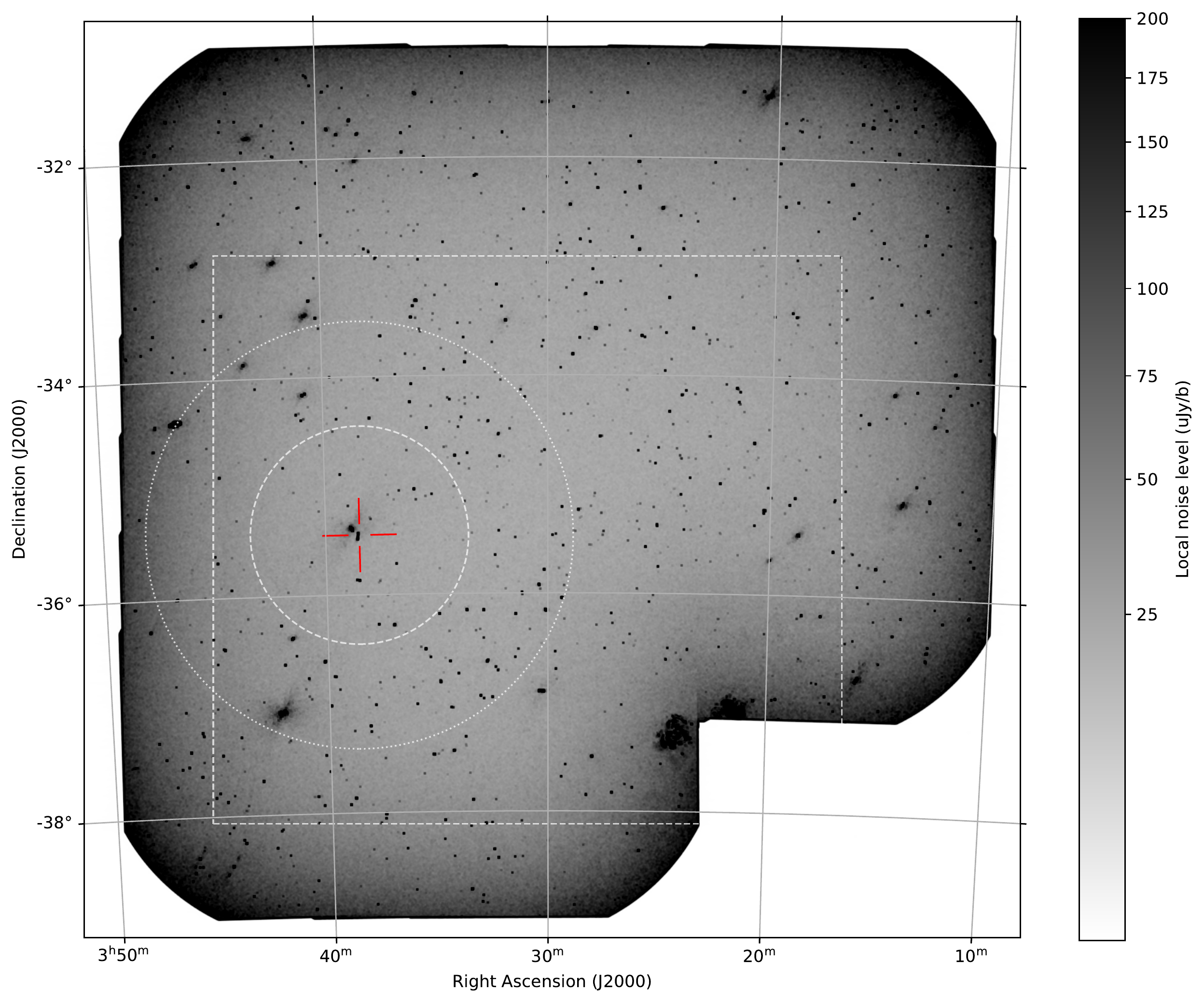}
\caption{The local root-mean-squared (RMS) noise in the peak-$P$ map. This is supplied in lieu of the peak-$P$ map itself, which renders point sources effectively invisible for our high resolution, large area map. This RMS map was generated by running a square sliding window of width and height both equal to five synthesised beamwidths over the peak-$P$ map, and calculating the RMS values of the pixels inside the window. The image shown here has a square root stretch applied. Linearly polarised radio sources are visible as a marked increase in the local RMS value. In source-free regions, the RMS is typically $\sim30$ $\mu$Jy beam$^{-1}$, except at the mosaic edges, and in the vicinity of bright sources, where the faint imprint of the synthesised beam manifests as narrow, diagonal fan--like structures. The centre of the Fornax cluster is indicated with a red cross-hair. Fornax A is partially visible in the bottom-right corner of the map, where six beams are missing due to beamforming errors. The white dashed box approximately indicates the region shown in Figure \ref{fig:fornax_voronoi}. The white dashed line indicates an angular radius of one degree, while the white dotted line indicates the 705 kpc (1.96 degree) virial radius of the cluster.}
\label{fig:peak_P_noise}
\end{figure*}

\begin{figure*}
\centering
\includegraphics[width=\textwidth]{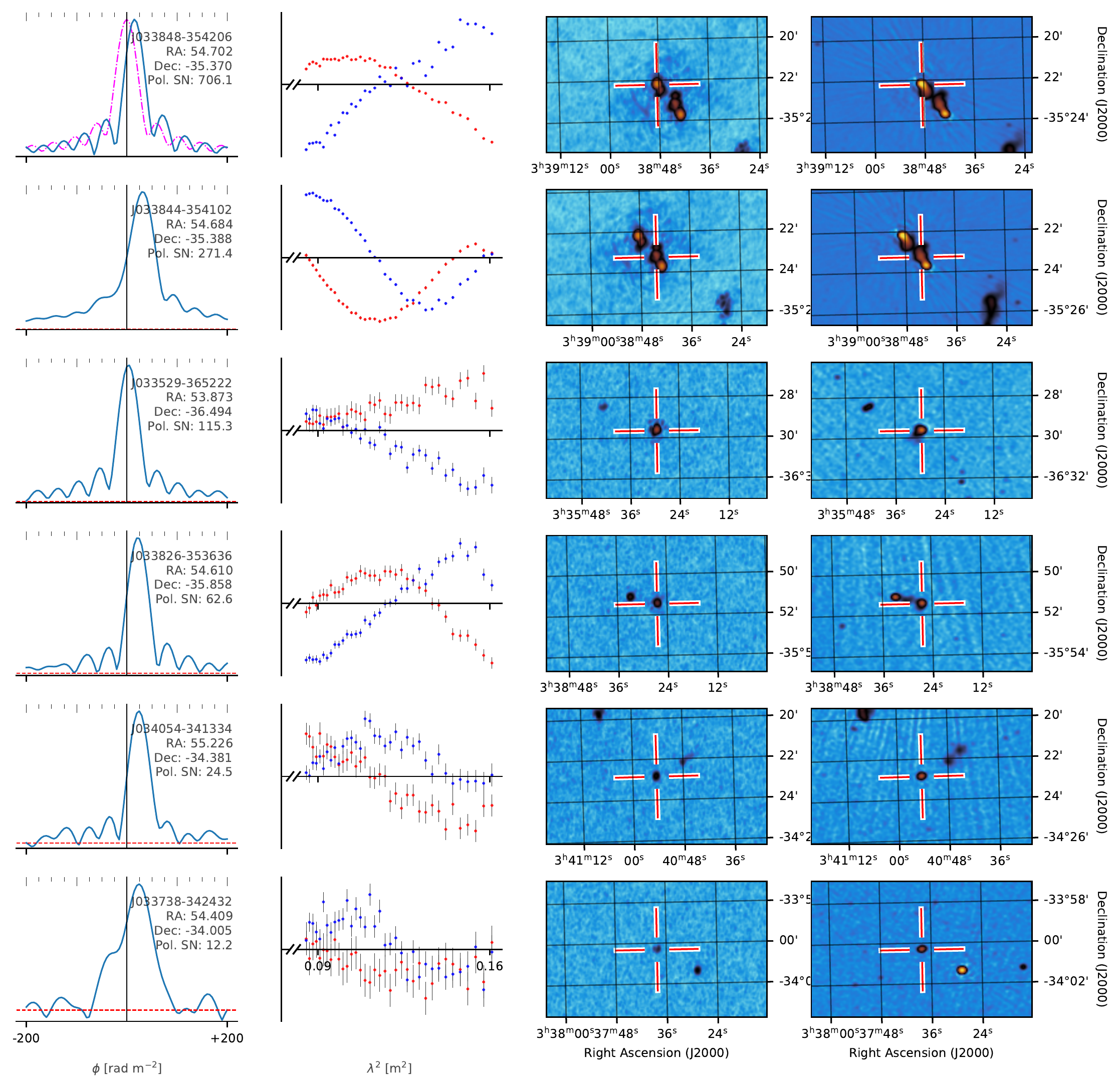}
\caption{The calculated dirty (i.e. no {\tt rmclean} \citep{Heald2009} performed; see Section \ref{sec:pol_analysis}) FDS (first column), corresponding Stokes $Q$ (red) and $U$ (blue) spectra (second column), peak-$P$ image (third column), and total intensity image (fourth column)}, for selected sources showing a range of polarised signal-to-noise. For columns 1 \& 2, the horizontal axes range from -200 to +200 rad m$^{-2}$ for the FDS plots (first column), and 0.08 to 0.16 m$^2$ for the Stokes ($Q$,$U$) plots (second column); note that tick labels are included on the bottom-most horizontal axes only. The vertical axes limits are all scaled to the maximum amplitude of the data points in individual plots. The J2000 name, right ascension, declination, and band-averaged polarised signal-to-noise ratio (SN) are all written in the respective FDS plots. The error bars on the ($Q$,$U$) data points indicate the standard deviation measured per image channel from the Stokes ($Q$,$U$) datacubes in a small region adjacent to each source. The peak polarised intensity of the sources are (from left to right, top to bottom) 212, 81, 3.5, 1.9, 0.7, and 0.4 mJy/beam/RMSF. Note that because the FDS have not been deconvolved with {\tt rmclean}, the emission-free regions of the FDS cannot be used as a guide to the underlying noise level. The RMSF (which is common to all of our sources, given our method) is plotted as a magenta dot-dashed line in the top-most FDS plot, scaled to the magnitude of the accompanying FDS. Note that the bottom-most source is a possible example of a source with multiple FDS emission peaks, but the $6\sigma$ S/N of the secondary peak is barely significant due to polarisation bias \citep{Macquart2012, Hales2012}. The peak-$P$ and total intensity images presented in columns 3 \& 4 each span $6.9\times11.3$ arcminutes, and are presented with a logarithmic image scaling. The red and white cross-hairs indicate the position at which the polarised data were extracted for the source in question. The first two rows illustrate how we have sampled independent lines of sight towards a resolved radio source; see discussion in Section \ref{sec:pol_analysis}.
\label{fig:selected_FDS_QU_plots}
\end{figure*}

We generated a map of the peak polarised intensity (peak-$P$) across the field from the FDS cube using {\tt Miriad}'s \citep{Sault1995} {\sc moment} function (see Figure \ref{fig:peak_P_noise}, which shows the associated RMS noise map, since the resolution and sky coverage of our observations would render most sources invisible in the peak-$P$ map itself). We then identified polarised radio sources in the field by applying the {\tt Aegean Software Tools} source-finding package \footnote{\url{https://github.com/PaulHancock/Aegean}, version 2.0.2} \citep{Hancock2012,Hancock2018} to our multi-frequency synthesis (MFS) Stokes $I$ map and the peak-$P$ map, using seedclip [floodclip] values\footnote{The {\sc Aegean} algorithm incorporates the generic {\sc Floodfill} algorithm. In an image, source detections are `seeded' at all locations where the pixel values exceeds the seedclip value. The algorithm then walks outwards from these locations, incorporating all adjacent pixels that exceed the floodclip value into the source model. Details are supplied in \citet{Hancock2012}.} of 5 [4] and 7 [5] respectively. With these settings, the nominal false detection rate (FDR) expected from {\sc Aegean} is $\sim2\%$ and $\sim0.5\%$ for the total intensity and peak-$P$ images respectively, though we note that the true FDR will be somewhat higher because (a) the peak-$P$ image contains complicated non-Gaussian noise structure (e.g. see \citealp{Hales2012}), which has the effect of pushing noise peaks to higher apparent significance under the assumption of Gaussian noise statistics, and (b) the peak-$P$ and total intensity images each contain residual deconvolution artifacts around bright sources. We ameliorated these relatively high effective FDRs by cross-matching the source-finding results from the peak-$P$ and total intensity images. Assuming that the peak-$P$ and total intensity images and associated source finding results are statistically independent, the resulting nominal baseline FDR is a negligible $\sim0.01$\%. Most importantly though, the cross-matching provided almost total suppression of false detections from deconvolution artifacts near bright sources, as determined by careful visual examination of the results. Our source-finding parameters yield a $7\sigma$ signal-to-noise cut in band-averaged linear polarisation, which is required for reliable RM measurements (e.g. \citealp{Macquart2012}). 

For this background RM grid experiment, we extracted the peak polarised intensity (peak-$P$) and associated sensitivity (see Figure \ref{fig:peak_P_noise}), and the peak Faraday depth of the source ($\phi_\mathrm{peak}$), from the dominant peak in the Faraday spectrum from each source. In practice, almost all of sources only had a single peak in the FDS. The quality of the FDS (and the associated Stokes $Q$ and $U$ spectra) are good throughout the mosaic; representative examples of FDS and their associated (Q,U) vs. $\lambda^2$ spectra are shown in Figure \ref{fig:selected_FDS_QU_plots}, which were selected at random to span the range of polarised signal-to-noise of the sources included in our RM grid sample. The full catalogue is provided online\footnote{<insert web address here>}.

Since the vast majority of sources detected were spatially unresolved or nearly so, we extracted the Stokes $I$, peak-$P$, and $\phi_\mathrm{peak}$ values of each source at the location of the brightest pixel in the peak-$P$ map. For the few heavily-resolved sources in the map (i.e. PKS B0336--35, which is actually comprised of the radio source inside NGC 1399, and a physically un-associated source several arcminutes to the NE --- see \citealp{Killeen1988}), we extracted the aforementioned quantities at the central coordinate location of the Gaussian emission components comprising the islands outputted by {\tt Aegean}. This results in samples of a suitable number of independent lines of sight towards these resolved sources. 
The polarisation state of some of these sources will be dominated by off-axis polarisation leakage. Since a robust, frequency-dependent, off-axis polarisation calibration procedure has not been finalised for ASKAP, we proceeded by identifying and eliminating such sources from our sample. We estimated the position-dependent, frequency-\emph{independent}, Stokes $I\rightarrow Q$ and $I\rightarrow U$ leakages using field sources, as described in Appendix \ref{sec:appendixa}. The results are that the leakages are lower than 1\% in most of the mosaic, but are greater in some areas, and in particular, the mosaic edges and corners (see Appendix \ref{sec:appendixa}). We eliminated sources from our sample that: 

\begin{enumerate} 
\item were located more than $3.5^\circ$ from the mosaic centre
\item were not located inside the half power point (at 1027 MHz) of at least one formed beam  
\item had measured fractional Stokes $q,u$ (we define $q=Q/I$, $u=U/I$, and use this nomenclature henceforth) values within $3\sigma$ uncertainty of our leakage map predictions at that location
\item had measured Stokes $q$ and $u$ values that were simultaneously within a factor of 2 of our leakage map predictions at that location
\end{enumerate} 

\begin{figure}
\centering
\includegraphics[width=0.45\textwidth]{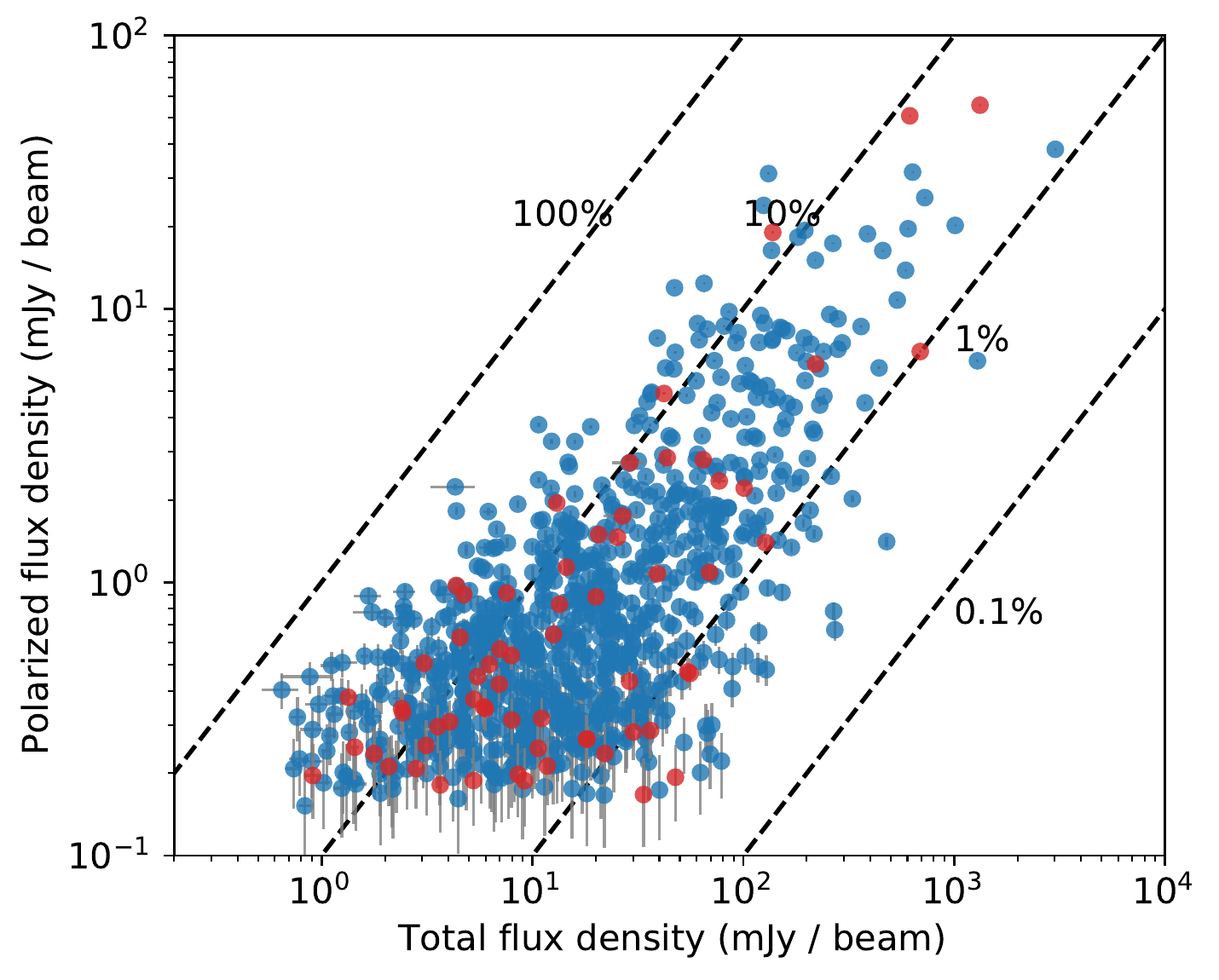}
\caption{Linearly polarised (un-debiased) vs. total flux density for the 870 sources in our sample. The red points represent sources inside a projected cluster-centric distance of 1 degree, while the blue points represent the converse. This distinction becomes relevant in Section \ref{sec:trend}. The dashed diagonal lines are lines of constant fractional polarisation (from top left to bottom right: 100\%, 10\%, 1\%, 0.1\%).}
\label{fig:fornax_source_flux_dist}
\end{figure}

In combination, the first two criteria ensure that sources are observed close to at least one beam centre, that multiple beams contribute to the final mosaic at the source locations, and that regions of high $I\rightarrow U$ leakage found outside the centres of the corner beams in the {\tt square\_6x6} beam footprint are excluded from our analysis (see Appendix \ref{sec:appendixa}, Figure \ref{fig:leakage_higherrez_zoom_combined}). In turn, this ensures that the off-axis response is averaged down, and that the polarisation of the sources are measured in multiple beams that can be evaluated for consistency. The latter two criteria (respectively) ensure that the polarisation state of a source is not either (a) consistent with pure instrumental leakage, or (b) dominated by instrumental leakage. We note that we also tested other methods to exclude spurious leakage-dominated sources, including local cuts on fractional polarisation based on predictions from our leakage maps, and uniform cuts on sources with fractional polarisations as high as 1.5\%. Our results were not significantly affected by the choice of method. After the cuts listed above, our sample consists of 870 linearly polarised sources, with a median fractional polarisation of 4.8\% (uncorrected for Ricean polarisation bias; see e.g. \citealp{Hales2012}, and noting the additional upward bias on the sample median fractional polarisation imposed by our polarised intensity cutoff; see Figure \ref{fig:fornax_source_flux_dist}). Figure \ref{fig:fornax_source_flux_dist} plots the distribution of polarised vs. total intensity for this sample, which appear broadly consistent with distributions derived from several similar $\sim$GHz-frequency studies \citep{Feain2009,Banfield2014,Hales2014}. Moreover, the sample yields an average polarised source density of 27 per square degree. This is consistent with several predictions for the number density of linearly polarised sources for GHz-frequency surveys with similar depth and resolution \citep{Stil2014,RO2014,RO2014e}, and is somewhat \emph{lower} than others (e.g. \citealp{Hales2014}). We are therefore confident that spurious leakage-dominated sources do not significantly contaminate our sample, and do not affect our analysis or conclusions in any important way.

The Galactic contribution to Faraday rotation measure is of order 10 rad m$^{-2}$ in this region (refer to Table 1 for Galactic coordinates), but varies over the field (\citealp{Anderson2015}, and see below). We attempted to remove this foreground contribution by fitting and subtracting a 2nd degree polynomial surface to the position-dependent $\phi_\mathrm{peak}$ values of our sources to yield $\phi_{\mathrm{peak,res}}$ --- the residual peak Faraday depth. Sources located within 1.5 degrees projected distance of the Fornax A radio core were excluded from the fit. We tested alternative fitting approaches, such as using a planar surface instead of the 2nd degree polynomial surface, and both the planar and 2nd degree polynomial surfaces after excluding data points located within 1.5 degrees of the cluster centre (the reason for which will become apparent in Section \ref{sec:rms}). In all cases, the residual RMs did not differ substantially in the vicinity of the cluster or throughout the larger field. Denoting the right ascension and declination a given source in decimal degrees in the J2000 epoch as $x$ and $y$, we define a 2nd degree polynomial surface as $p(x,y)=\sum_{i,j} c_{i,j}*x^i*y^j$ with $i,j\leq2$, then  fitting over the field as described above, we derive best-fit model coefficients of $c_{0,0}=-4.88001154\times10^5$, $c_{0,1}=-2.76553274\times10^4$, $c_{0,2}=-3.91350846\times10^2$, $c_{1,0}=1.84508441\times10^4$, $c_{1,1}=1.04492833\times10^3$, $c_{1,2}=1.47768421\times10^1$, $c_{2,0}=-1.74583394\times10^2$, $c_{2,1}=-9.88225943$, $c_{2,2}=-0.139675724$. Over the field, this model is essentially consistent with the all-sky Galactic RM model derived by \citet{HE2020}, both qualitatively and quantitatively. In both cases, the model has a mean value of $\sim+10$ rad m$^{-2}$, showing a slight gradient running almost directly N-S through the field, from $\sim+20$ rad m$^{-2}$ for the northernmost sources down to $\sim+3$ rad m$^{-2}$ for the southernmost sources. Our model is systematically $\sim5$ rad m$^{-2}$ lower than the \citet{HE2020} model in the very northern-most part of our field, but the discrepancy is generally less than or equal to the \citet{HE2020} model uncertainty in this region, and we have verified that the difference does not affect our results or conclusions in any case.

The uncertainty in $\phi_{\mathrm{peak,res}}$ was calculated as per \citet{BdB2005}, based on the peak-$P$ value of each source, the RMS noise measured from band-averaged maps of Stokes $Q$ and $U$ in an adjacent source-free region (typically 30 $\mu$Jy beam$^{-1}$ per Stokes parameter), and the rotation measure spread function (RMSF, which is the point-spread function in Faraday depth space; see \citealp{BdB2005}) width measured directly from our data accounting for our channelisation scheme. We have multiplied the uncertainties in $\phi_{\mathrm{peak,res}}$ by an additional factor of 1.2, to capture the aggregate effect of uncorrected widefield polarisation leakage (see Section 5.2 of \citealp{Ma2019}).

\section{Ancillary analysis}\label{sec:ancillary}

\subsection{Cluster-relative line-of-sight source positions from redshift catalogue cross-matching}\label{sec:redshifts}

The vast majority of radio sources brighter than $\sim1$mJy at $\sim$GHz-frequencies are powerful AGN that lie at far greater distance than the Fornax cluster (e.g. \citealp{Magliocchetti2000,GW2008,deZotti2010}). Nevertheless, it is desirable to confirm this for the sources used in our particular RM grid experiment, for reasons described in Section \ref{sec:intro}. 

\citet{Maddox2019} have compiled a catalogue of reliable spectroscopic redshifts towards the Fornax cluster, drawn from deep optical imaging surveys in the literature as well as their own data. This catalogue is complete within a degree of NGC 1399, down to brightness levels typical of faint ultra-compact dwarf galaxies and globular clusters. The objects in our sample are all brighter than 0.6 mJy/beam in total radio intensity (see Figure \ref{fig:fornax_source_flux_dist}) and will, if located in the Fornax cluster redshift range ($600<cz<3000$ km s$^{-1}$), have optical counterparts at least as bright as a star-forming galaxy (e.g. \citealp{Padovani2016}). It follows that any of our sample sources inside the Fornax cluster will have a spectroscopic redshift in the Maddox \emph{et al.} catalogue.

We cross-matched our sample against the Maddox \emph{et al.} catalogue for objects which (1) lie within 1 degree of the centre of NGC 1399 (this radius becomes relevant in Section \ref{sec:trend}), (2) had redshifts consistent with lying inside the Fornax cluster volume, and (3) were not associated with `Galactic stars' or `globular clusters' in Maddox \emph{et al.}. We used an initial matching radius of 90 arcseconds to account for possible offsets between steep spectrum double radio sources and their optical counterparts (e.g. \citealp{Hammond2012}). This produced three candidate matches, apart from the radio source associated with NGC 1399 itself. However, each candidate was then found to be a single, isolated, spatially-unresolved radio source. For such sources, a matching radius equal to the synthesised primary beam width (10 arcseconds) is more appropriate, but even a 30 arcsecond matching radius eliminates all of the initial candidate matches. This complete lack of optical redshift counterparts confirms that our entire sample (with the exception of the radio source hosted by NGC 1399) lies beyond the Fornax cluster, and that their RMs are accumulated along lines-of-sight that traverse the entire distance through the cluster. 

Relaxing the cluster volume redshift range constraint, but otherwise cross-matching using the same methods against the full Maddox \emph{et al.}. catalogue, we obtain 21 matches satisfying $0.0048<z<2.41$, with $<z>=0.65$. Six of these sources are located at a projected distance of less than one degree, and have $0.197<z<1.606$ and $<z>=0.902$. Outside one degree, there is one source inside the cluster redshift range with $z=0.0048$. The remainder of this sub-sample satisfies $0.079<z<2.41$ and $<z>=0.55$. Therefore, to the extent possible, we confirm that our sample sources typically reside well beyond the Fornax cluster, though somewhat closer than is typical for moderately bright (>10 mJy/beam) radio sources in the NRAO VLA Sky Survey (NVSS, for which $<z>\approx1.2$; see \citealp{Brookes2008} and Figure 11 of \citealp{deZotti2010}). This is not unexpected, given the complex differences in strategy used by the redshift surveys involved. Finally, a 2-sample Kolmogorov-Smirnov test shows that the redshift distributions of cross-matched sources located inside versus outside 1 degree do not differ significantly from each other (D-value = 0.25, p-value = 0.89). Again, this scale becomes relevant in Section \ref{sec:trend}. We also looked for any significant dependence of RM on redshift, but given the small size of the sample, we were unable to provide any reasonable constraints.

\section{Results}

\subsection{Faraday depths are enhanced near the cluster}\label{sec:rms}

\subsubsection{Global trend in Faraday depth versus projected distance}\label{sec:trend}

\begin{figure*}
\centering
\includegraphics[width=\textwidth]{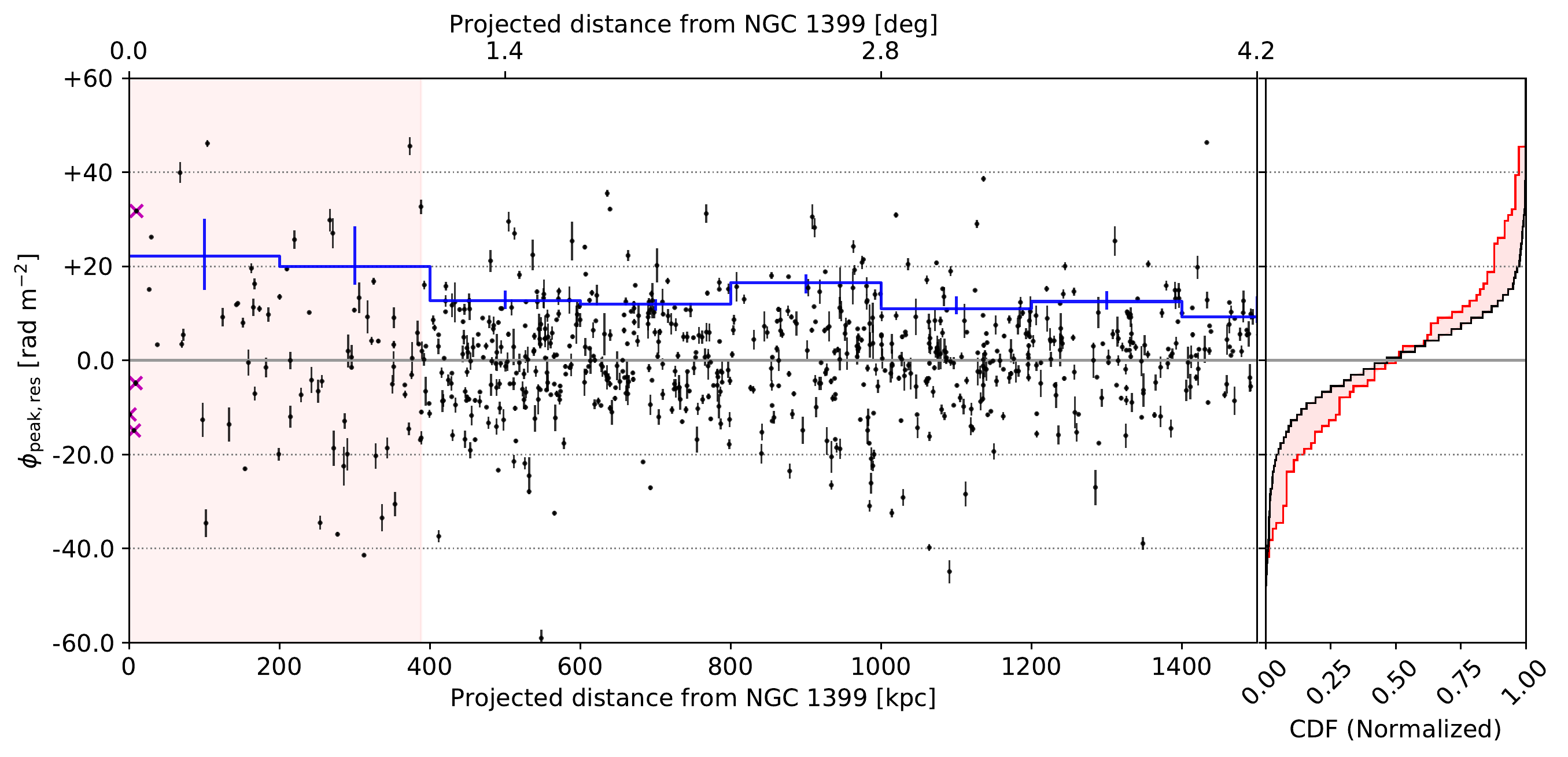}
\caption{Foreground-corrected Faraday depth ($\phi_{\mathrm{peak,res}}$) versus projected distance from NGC 1399. The foreground was removed as described in Section \ref{sec:rms}. Data points within one degree (indicated by the red shaded region) show an excess dispersion, as described in Section \ref{sec:rms}. Sources that are located inside the Fornax cluster volume, instead of behind it, are indicated with magenta crosses (see Section \ref{sec:ancillary}). Note that all such sources are in fact sub-components of the central radio source in NGC 1399 (following from our approach for dealing with heavily resolved sources, discussed in Section \ref{sec:pol_analysis}).} The blue step plot shows the half-interdecile range (i.e. the interdecile range divided by two) for the data points located within each 0.56 degree (200 kpc) -wide step. The vertical blue bars indicate the associated 90\% confidence interval for the underlying population distribution in each bin, calculated using bootstrap re-sampling. The right-most axes show normalised cumulative histograms of $\phi_\mathrm{peak}$ for sources located within (red) and outside (black) a projected distance of one degree. The red shading highlights the difference between these distributions.
\label{fig:peak_FD_vs_impact_param}
\end{figure*}

Figure \ref{fig:peak_FD_vs_impact_param} plots $\phi_{\mathrm{peak,res}}$ as a function of distance from the centre of the cD-type galaxy NGC 1399, which we take to be the position of the centre of the cluster (\citealp{Drinkwater2001}; i.e. located at $03^h38^m28.68^s$, $-35^d27^m07.14^s$ (J2000), and indicated on Figure \ref{fig:peak_P_noise} with a red cross-hair). Note that our sample selection criteria (with respect to the mosaic beam centres; see Section \ref{sec:pol_analysis}) mean that sources are truncated eastward of RA = 03h45m, 1.33 degrees east of the cluster centre at the same declination.

\begin{figure}
\centering
\includegraphics[width=0.43\textwidth]{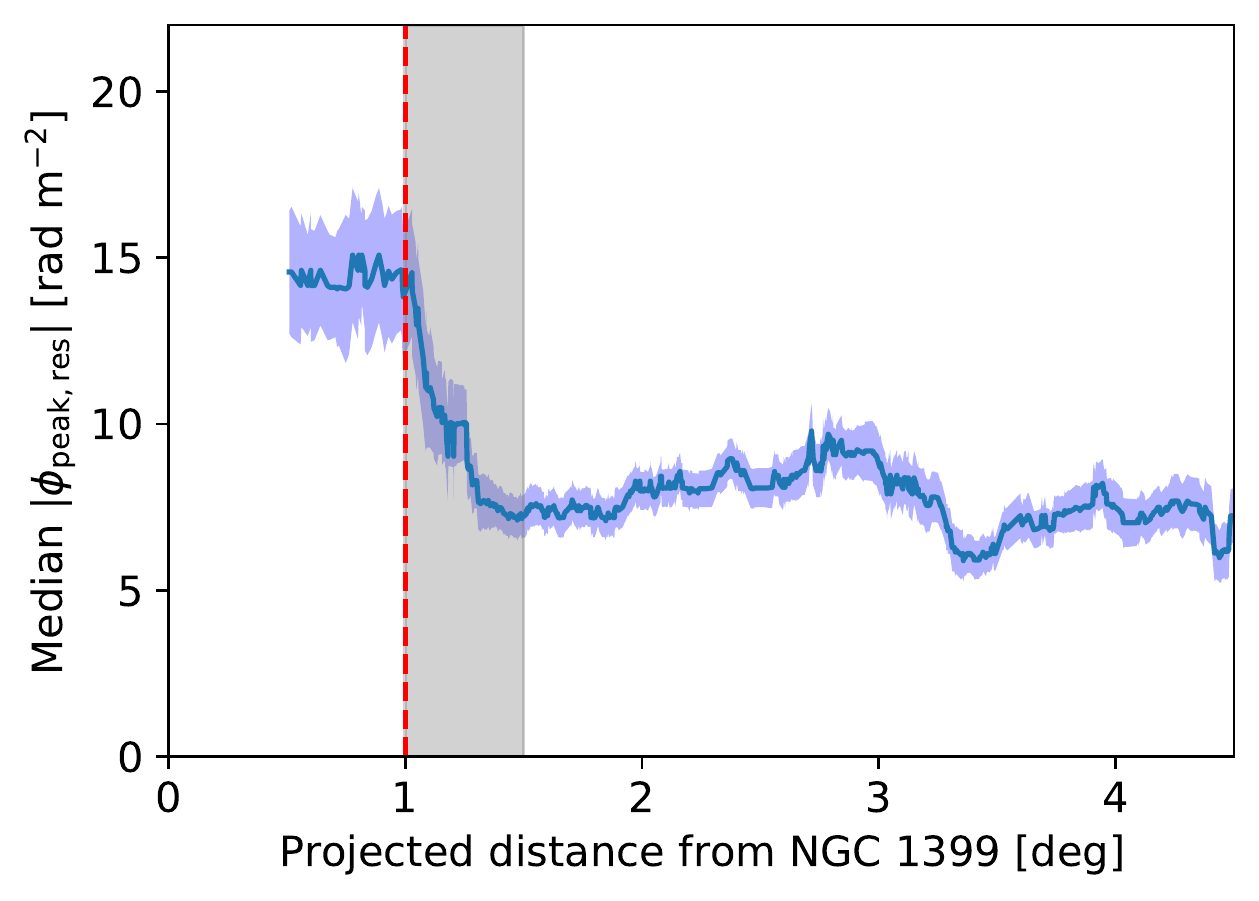}
\caption{The median of |$\phi_{\mathrm{peak,res}}$| in a sliding window of width 0.5 degrees as a function of the cluster-centric radius of the outer bound of this window (blue line). The blue-shaded region indicates the 95\% confidence interval on this value, calculated as $\pm1.58\times\text{IQR}/\sqrt{n}$ \citep{McGill1978}, where IQR and $n$ are the interquartile range and number of measurements (respectively) of |$\phi_{\mathrm{peak,res}}$| in the sliding window. A sharp and significant decrease in the plotted values is evident when the outer bound of the window passes a cluster-centric radius of 1 degree, which is marked with a vertical red dashed line. The width of the sliding window is indicated by the gray shaded region.}
\label{fig:fornax_cluster_sliding-median_abs_RM}
\end{figure}

The left-hand panel of Figure \ref{fig:peak_FD_vs_impact_param} shows an increase in the dispersion of $\phi_{\mathrm{peak,res}}$ within $\sim1$ degree (360 kpc) of the cluster centre ($\sigma=20.5$ rad m$^{-2}$), compared to sources outside this radius ($\sigma=11.8$ rad m$^{-2}$). Figure \ref{fig:fornax_cluster_sliding-median_abs_RM} shows a similar result, but with finer granularity. Here, we have calculated the median of |$\phi_{\mathrm{peak,res}}$| in a sliding window of maximum width 0.5 degrees as a function of the cluster-centric radius of the outer bound of this window. A sharp decrease in the median of |$\phi_{\mathrm{peak,res}}$| is clearly evident as the outer bound of the sliding window crosses 1 degree, and has dropped to a more-or-less constant lower value at 1.5 degrees, indicating that a sharp transition in the degree of observed Faraday rotation occurs at the former distance.

The difference in Faraday rotation inside vs. outside 1 degree cannot be accounted for by the measurement uncertainties. A 2-sample K-S test applied to the cumulative distributions for these sub-samples (also shown in Figure \ref{fig:peak_FD_vs_impact_param}) confirms that they differ both substantially and significantly, yielding D- and p-values of 0.32 and $3\times10^{-7}$ respectively. The sharp break in $\phi_{\mathrm{peak,res}}$ dispersion, and the relative enhancement in this dispersion towards smaller cluster-centric radii, persist regardless of how the data are binned or the dispersion is parameterised --- see for example the half-interdecile range plot calculated in 0.56 degree (200 kpc) -wide bins in Figure \ref{fig:peak_FD_vs_impact_param}. Thus, we take the mean Faraday depth contribution of the cluster plasma to be $\sigma_{\phi_{\mathrm{peak,res,cluster}}}=\sqrt{20.5^2-11.8^2}=16.8\pm2.4$ rad m$^{-2}$, where the quoted uncertainty range corresponds to the 95\% confidence interval calculated via bootstrap re-sampling. 

Apart from their difference in $\phi_{\mathrm{peak,res}}$ dispersion, the distributions are otherwise similar --- the skew and excess kurtosis of both do not significantly differ from the values expected for the Normal distribution (i.e. zero), for example.  

\begin{figure*}
\centering
\includegraphics[width=0.95\textwidth]{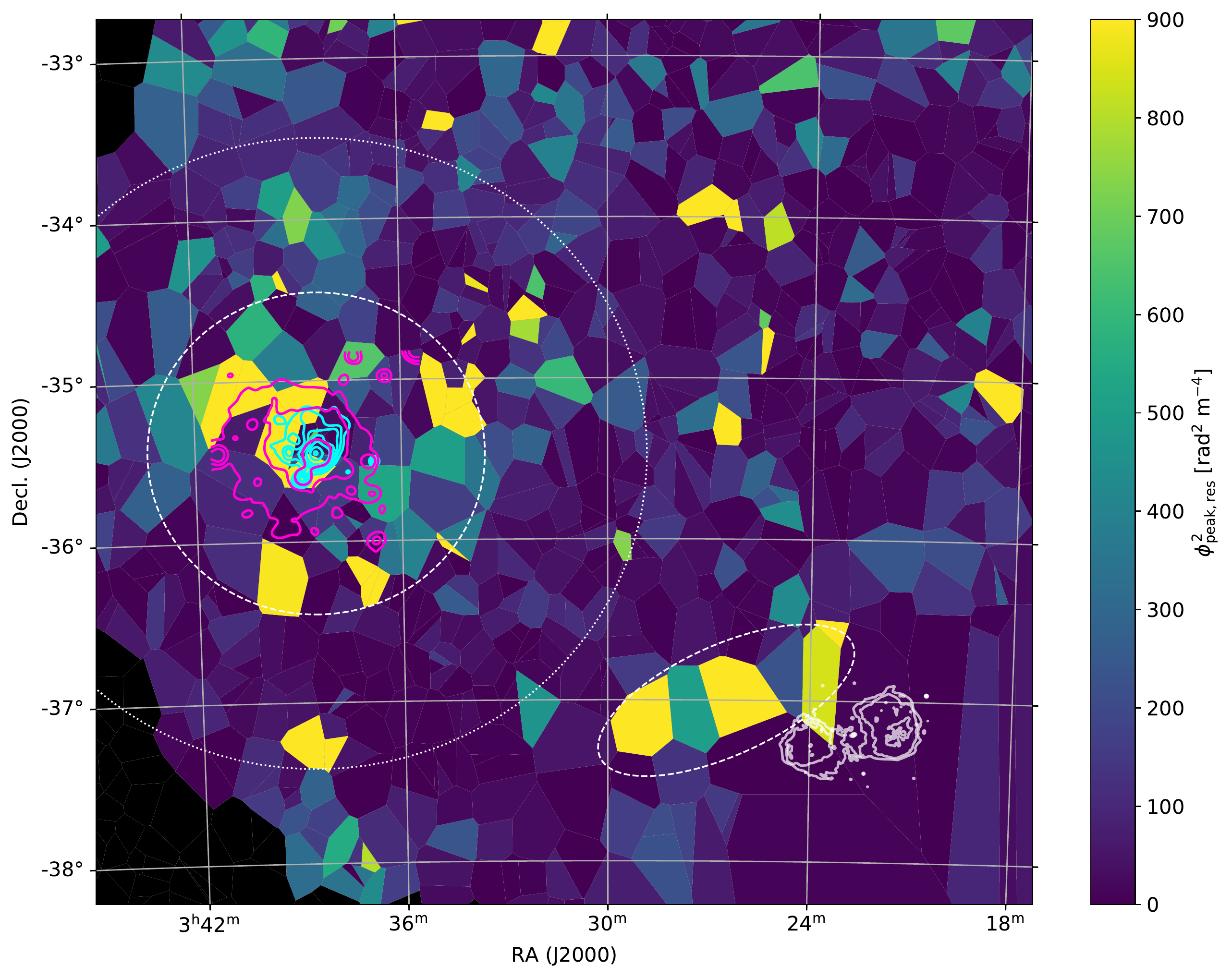}
\caption{A map of $\phi_{\mathrm{peak,res}}^2$ across the field, employing nearest neighbour interpolation, as described in Section \ref{sec:rms}. The region shown is indicated in its broader context in Figure \ref{fig:peak_P_noise} with a white dashed box. Each cell contains a single polarised source, and is colorised by the sources' value of $\phi_{\mathrm{peak,res}}^2$. The extent of X-ray emission from the Fornax cluster ICM as seen by \emph{Chandra} (0.3–1.5 keV bandpass; light blue contours; smoothed to 2.5 arcminute resolution; \citealp{Scharf2005}) and the \emph{ROSAT} Position-Sensitive Proportional Counter (PSPC; 0.1—2.4 keV; smoothed to 3 arcminute resolution; pink contours; \citealp{Jones1997}) is indicated. White contours show Fornax A. The white dashed circle indicates 1$^\circ$ projected distance --- the projected distance inside which the variance of $\phi_{\mathrm{peak,res}}$ was found to be enhanced in Figure \ref{fig:peak_FD_vs_impact_param}. The white dashed ellipse roughly indicates where $\phi_{\mathrm{peak,res}}^2$ values appear to be elevated in a contiguous region near Fornax A. The white dotted line indicates the 1.96 degree (705 kpc) virial radius of the cluster \citep{Iodice2017}. The blacked-out polygons indicate sources which fall more than $3.5^\circ$ from the mosaic centre, and which are therefore excluded from our polarimetric analysis (Section \ref{sec:pol_analysis}).}
\label{fig:fornax_voronoi}
\end{figure*}

\subsubsection{Spatial morphology of the Faraday depth enhancements}\label{sec:spatial}

The morphology of the $\phi_{\mathrm{peak,res}}$ enhancement is revealed in Figure \ref{fig:fornax_voronoi}, where we interpolate (using the nearest neighbour method) and plot $\phi_{\mathrm{peak,res}}^2$ as a function of position. Contours showing the observable extent of the X-ray-emitting ICM as seen by \emph{Chandra} \citep{Scharf2005} and \emph{ROSAT} \citep{Jones1997} are overlaid, as are contours for the radio galaxy Fornax A (see Norris \emph{et al. submitted}). We note that the \emph{Chandra} and \emph{ROSAT} X-ray maps are smoothed to 2.5 and 3 arcminutes respectively to reveal the faint diffuse emission. Particularly for the \emph{ROSAT} contours, the apparent degree of extension towards the southwest is affected by the smoothing required combined with the presence of bright point sources unrelated to the cluster medium. The ICM is therefore more asymmetric around NGC 1399 --- more `swept back' towards northeast --- than the \emph{ROSAT} contours initially seem to imply (echoing the morphology of the \emph{Chandra} contours). The dispersion in $\phi_{\mathrm{peak,res}}$ is clearly enhanced in the vicinity of the main cluster, extending to a radius of one degree in most directions, and to 1.5--2 degrees towards the N and NW. The radial extent of the enhanced region exceeds that of the currently observable X-ray emitting ICM by a factor of 2--4 depending on azimuthal bearing, but lies within the 1.96 degree (705 kpc) virial radius of the cluster (\citealp{Iodice2017}; indicated on Fig. \ref{fig:fornax_voronoi}). 

\begin{figure*}
\centering
\includegraphics[width=\textwidth]{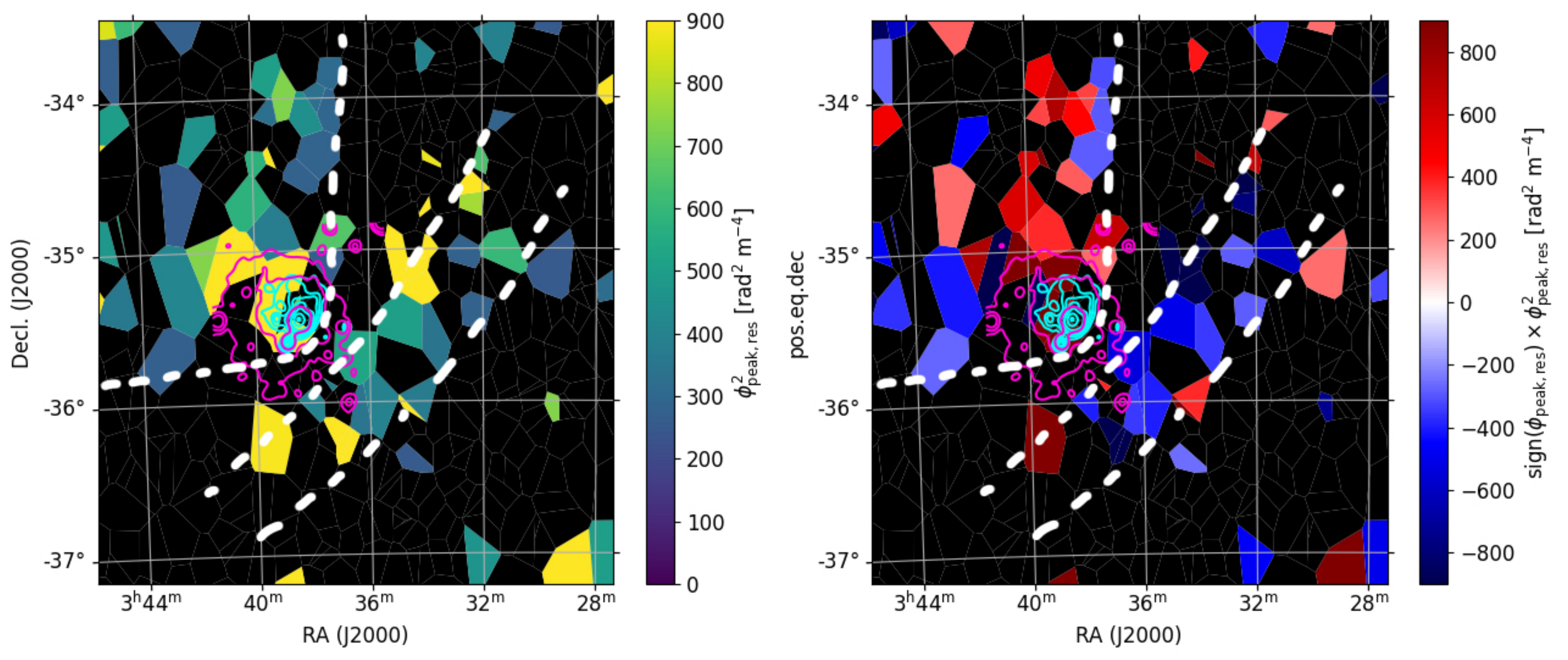}
\caption{\emph{Left:} As for Figure \ref{fig:fornax_voronoi}, but zoomed on the main Fornax cluster, and with $|\phi_{\mathrm{peak,res}}^2|<200$ rad$^2$m$^{-4}$ masked (appearing black). The two distinct regions of enhancement described in the main text are delineated by white dashed lines. \emph{Right:} As for the left panel, but showing $\text{sign}(\phi_{\mathrm{peak,res}})~\times~\phi_{\mathrm{peak,res}}^2$.}
\label{fig:scenario}
\end{figure*}

The global $\phi_{\mathrm{peak,res}}$ enhancement appears to be comprised of two smaller sub-regions: the first, a circular sector of angle $\sim90^\circ$, with its vertex located near NGC 1399, and its two enclosing radii oriented slightly clockwise of N-S and E-W (respectively), and the second region, a $\sim0.5^\circ$-wide strip centred $\sim0.75^\circ$ SW of NGC 1399, slightly concave towards it. The sharp decline in $\phi_{\mathrm{peak,res}}$ dispersion at 1 degree cluster-centric radius is particularly evident at the outer edge of this feature. This `two region' interpretation is bolstered when the sign of the $\phi_{\mathrm{peak,res}}$ values are considered (see Figure \ref{fig:scenario}). The SW enhancement shows predominantly negative $\phi_{\mathrm{peak,res}}$ values, indicating that the magnetic field in this structure is oriented predominantly away from the observer. Conversely, the NE enhancement does not reveal an obvious bias towards either positive or negative values, implying that the magnetic field in this region is more isotropic.

We note the existence of another possible region of enhanced $\phi_{\mathrm{peak,res}}$ dispersion, which runs adjacent to the Fornax A lobes to their NE, and which is not obviously attributable to calibration or image artifacts. While comprised of only six sources, the $\phi_{\mathrm{peak,res}}^2$ values are clearly elevated above the typical surrounding values for field sources --- values for four of the sources lie at $>3\sigma$, with two more at $>2\sigma$. The probability of this happening in a contiguous region is correspondingly small.  

\subsection{Flux densities of discrete sources are reduced near the cluster}\label{sec:linpol}

\begin{figure}
\centering
\includegraphics[width=0.47\textwidth]{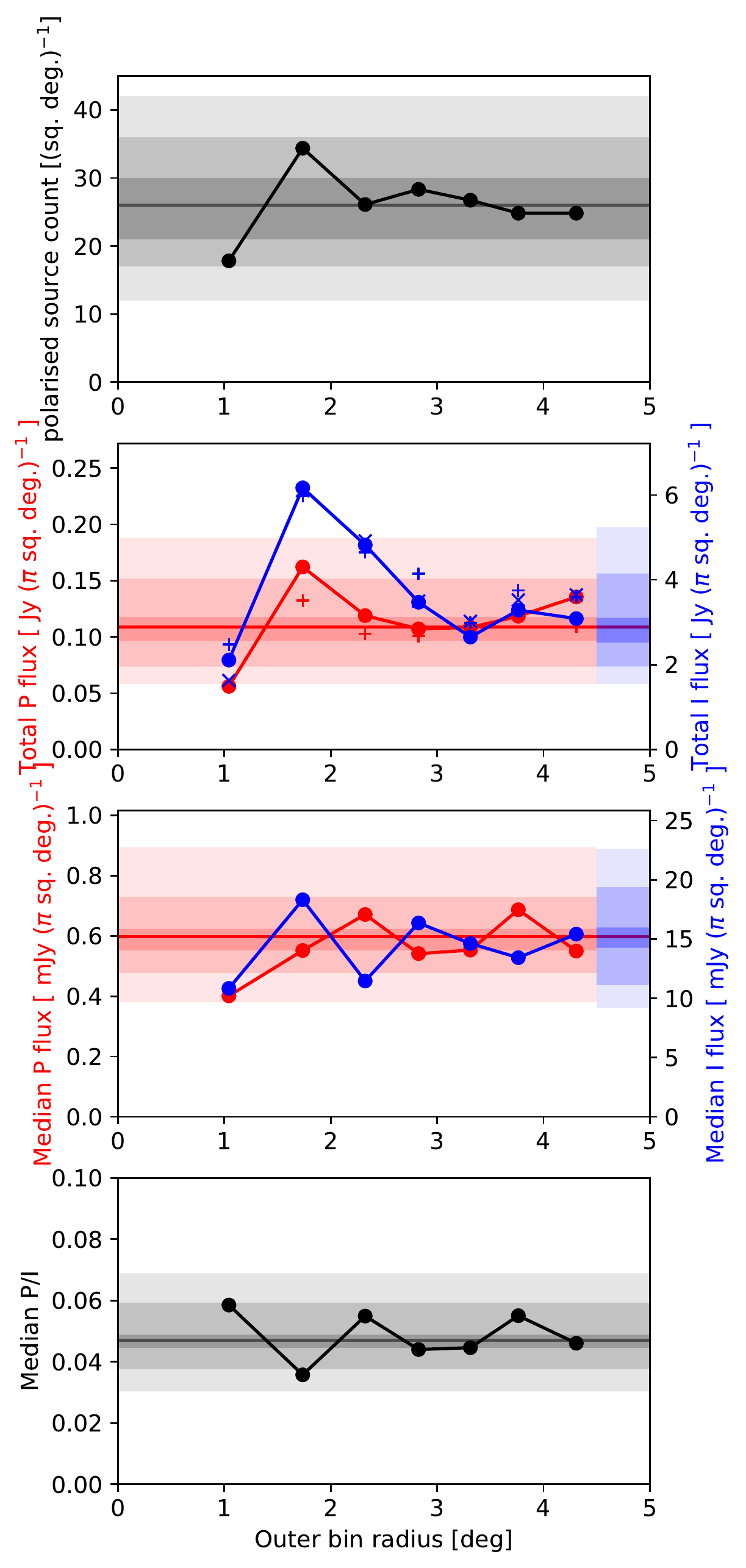}
\caption{Binned polarised source counts, polarised and total radio flux, and fractional polarisation statistics, calculated in equal $\pi$-square-degree annular bins centred on NGC 1399, and plotted against the bounding radius of each annulus. Details of the experiment are described in Appendix \ref{sec:appendixb}. \emph{Panel 1 (top):} The observed polarised source counts in each annular bin, expressed as the average number per square-degree. The grey bands indicate confidence intervals of 67\% (dark grey), 95\% (mid grey), and 99.7\% (light grey) around the expected count of 27 polarised sources per square degree (black horizontal line). \emph{Panel 2:} Integrated polarised (red) and total (blue) flux for ASKAP (joined dots), NVSS (`+' symbols) and GLEAM (`x' symbols; total intensity only) in each annular bin, with the NVSS and GLEAM fluxes scaled to those expected in the ASKAP frequency band by assuming a spectral index of $-0.7$. The fractional uncertainty on the plotted quantities is generally less than 1\%, and so are not indicated. The shaded areas indicate confidence intervals of 67\% (dark shading), 95\% (medium shading), and 99.7\% (light shading) for the ASKAP-derived polarised (red-shaded) and total intensity (blue-shaded) quantities, which have each been truncated horizontally in the plot for clarity. \emph{Panel 3:} As for panel 2, but here for the median flux in each annulus rather than its sum. \emph{Panel 4 (bottom):} The median fractional polarisation of sources in each annulus, calculated on a per-source basis. Confidence intervals are represented as above.}
\label{fig:counts_plus_flux_in_annular_bins_mod3}
\end{figure}

\begin{figure}
\centering
\includegraphics[width=0.47\textwidth]{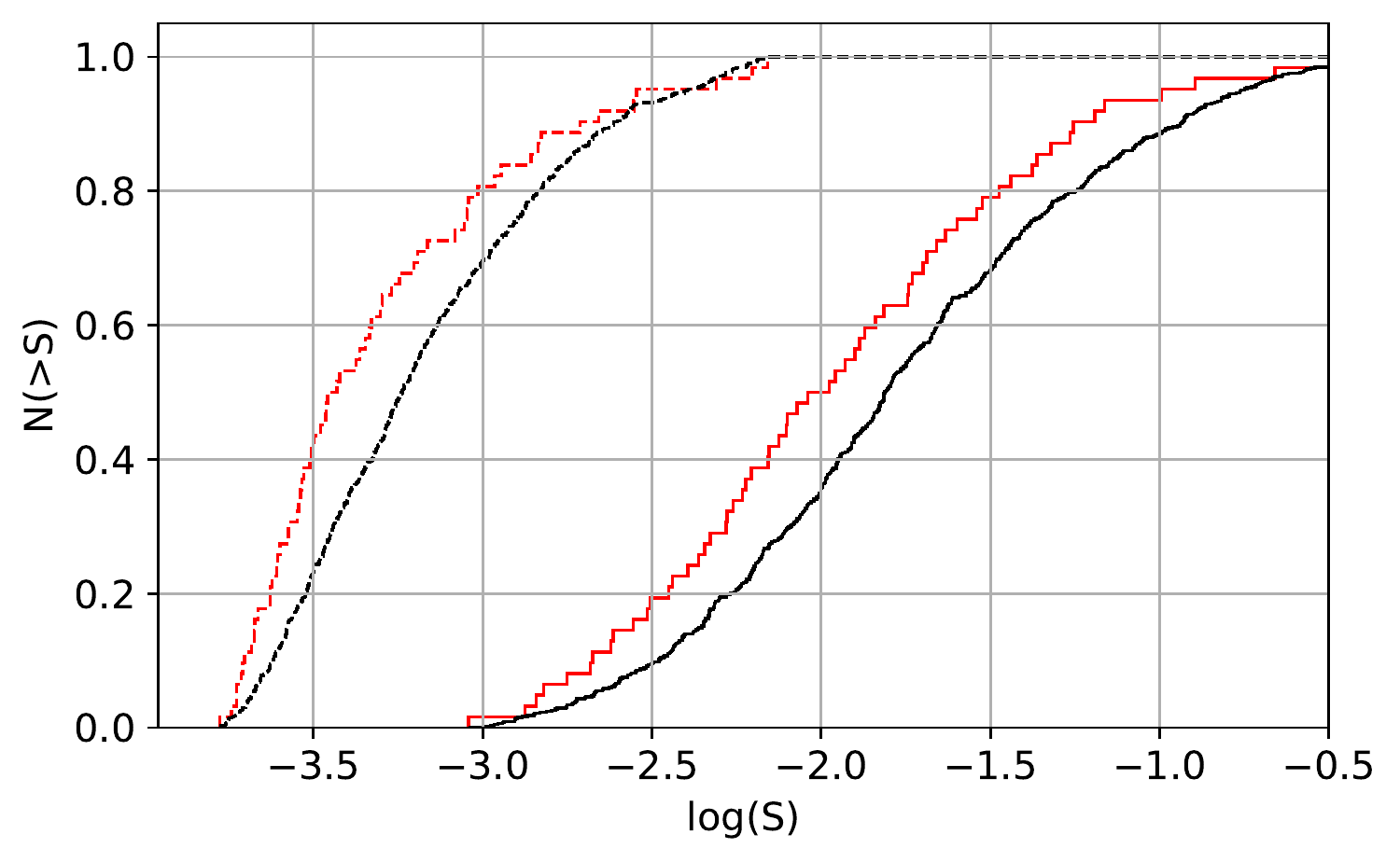}
\caption{Normalised integrated source counts versus the logarithm of polarised (dashed lines) and total (solid lines) flux density inside (red lines) and outside (black lines) 1 degree projected cluster-centric radius.}
\label{fig:source_counts}
\end{figure}

An apparent paucity of bright polarised sources in the vicinity of NGC 1399 (see the RMS map shown in Figure \ref{fig:peak_P_noise}) motivated us to consider whether small-scale structure in the cluster medium could be depolarising background sources (e.g. \citealp{Burn1966,Murgia2004,Bonafede2011}). A 2-sample Kolmogorov–Smirnov test comparing the polarised intensity distribution of our sample inside versus outside 1 degree results in a D-value of 0.22 with a p-value of 0.001, suggesting the lower apparent polarised flux near the cluster is statistically significant at the $\sim3.3\sigma$ level. The question is then whether sources near the cluster are fewer, fainter, or both, and whether depolarisation or some other mechanism is responsible for this.   

We investigated this by calculating the polarised source counts, and the integrated and median polarised and total flux, in equal-area ($\pi$-square degrees) annular bins centred on NGC 1399. The experiment is described in more detail in Appendix \ref{sec:appendixb}, while the results are shown in Figure \ref{fig:counts_plus_flux_in_annular_bins_mod3}. The polarised source counts (top panel of Figure \ref{fig:counts_plus_flux_in_annular_bins_mod3}) show a $2\sigma$ decrement inside one degree radius (averaging 17 polarised sources per square degree), a $2\sigma$ \emph{enhancement} between 1 and 2 degrees radius (averaging 34 polarised sources per square degree), but are consistent with our average 27 polarised sources per square degree thereafter. Taken on their own, these deviations are not statistically significant in a sky area the size of our mosaic. However, the plots of the integrated and median flux densities (middle and lower panels of Figure \ref{fig:counts_plus_flux_in_annular_bins_mod3}, respectively) mirror the behaviour of the source count plot, and here the deviations are significant. The integrated polarised and total fluxes (respectively) show $\sim3\sigma$ and $\sim2\sigma$ decrements relative to the expectation from the broader mosaic inside 1 degree radius, corresponding to a 50\% decrement in polarised flux in the former case. The data then show a $\sim2.5\sigma$ and $\sim4\sigma$ \emph{enhancement} in the integrated polarised and total flux (respectively) between 1 and 2 degrees radius. In total intensity only, the enhancement continues to fall outside the 99.7\% confidence interval until a radius of 2.3 degrees, beyond which it returns to low-significance deviations from the expectation value, along with the polarised flux. The median polarised and total fluxes also show $\sim2.5\sigma$ and $\sim3\sigma$ decrements (respectively) inside 1 degree radius, but less significant deviations in other bins. This may indicate that the flux decrement inside 1 degree is driven by the bulk of the sources therein, while the flux enhancement from 1--2.3 degrees may be driven by a smaller proportion of brighter sources. Finally, the median fractional polarisation shows no evidence for a decrement at small cluster-centric radii, as was found by \citet{Bonafede2011} for a sample of clusters, who attributed the effect to depolarisation by turbulent cells in the ICM as per our motivating hypothesis above. Instead, we find that the median fractional polarisation is $\sim5$\%, which is consistent with the values that \citet{Bonafede2011} derive at cluster-centric distances larger than 1.5 effective core radii (see figure 2 of that work). However, this comparison comes with the caveats that (a) our polarised sources at small cluster-centric radii is insufficient to probe the core ICM regions where \citet{Bonafede2011} observed the depolarisation effect, and (b) our RM grid sources are confirmed to lie exclusively behind the Fornax cluster, whereas \citet{Bonafede2011}'s sample was a heterogeneous mixture of sources embedded in the clusters being measured, as well as in the background (though see their Appendix A.1 for arguments that this cannot explain their results). In the future, it might be possible to probe closer to the core of the Fornax cluster using the central radio source hosted in NGC 1399 (e.g. \citealp{Killeen1988}), at which point a more detailed comparison would be appropriate.

We note that the decrement and enhancement structures described above are visually apparent in Figure \ref{fig:peak_P_noise} (in polarisation). The results are unusual for the Stokes $I$ emission in particular, for it is not obvious how emission or transmission processes linked to the cluster could produce them (see discussion in Section \ref{sec:discoreduced}). A mundane possibility is that, given the fact that the data were collected during ASKAP's Early Science phase, the effect could be instrumental. We rule this out conclusively by including data from the GaLactic and Extragalactic All-sky Murchison Widefield Array (GLEAM; 72--231 MHz; \citealp{Wayth2015,Hurley-Walker2017}) and NVSS (1.4 GHz; \citealp{Condon1998}) survey catalogues in the middle panel of Figure \ref{fig:counts_plus_flux_in_annular_bins_mod3}. We applied the same spatial truncations to these data as to our sample, and then scaled their fluxes to those expected in the ASKAP band assuming a spectral index of $-0.7$. Evidently, the modified total intensity data from GLEAM and NVSS track the ASKAP data closely, as does the polarised intensity data from NVSS, ruling out instrumental effects. A final question is whether these effects are observed over a range of flux densities. In Figure \ref{fig:source_counts}, we plot the normalised integrated source counts versus flux density, which we calculate both inside and outside 1 degree cluster-centric radius, for both polarised and total intensity. The data show that the decrement inside 1 degree radius persists over a wide range in flux density. 

Thus, we draw the following conclusions: (1) the polarised source counts, the polarised flux densities, and the total flux densities \emph{all} show decrements inside 1 degree projected radius from the cluster; (2) the average magnitude of the decrement is $\sim50\%$ for the polarised flux, and $\sim30\%$ for the total intensity; (3) between 1 and 2.3 degrees projected radius, there is a surfeit of flux in one or both quantities; (4) none of these results can be attributed to a normalisation problem, instrumental effect, or the modest $\sim50$\% increase in image noise that occurs in a small $\sim20\times10$ arcminute region around PKS B0336--35 (see Figure \ref{fig:peak_P_noise} and Section \ref{sec:pol_analysis}); (5) the coexistent decrements in Stokes $I$ and $P$ are inconsistent with our initial hypothesis that depolarisation by turbulent cells in the Fornax ICM could cause the decrement in $P$ (cf. \citealp{Bonafede2011}). We discuss these results further in Section \ref{sec:discoreduced}.

\subsection{Summary of observational results}\label{sec:results_summary}

We summarise our observational results as follows:

\begin{itemize}
    \item The distribution of peak Faraday depths for discrete radio sources shows an excess scatter of $\sigma_{\phi_{\mathrm{peak,res,cluster}}}=16.8\pm2.4$ rad m$^{-2}$ within 1 degree (360 kpc) of the Fornax cluster centre. In more spatially-limited areas, the excess can be traced out to the $\sim1.96$ degree (705 kpc) virial radius of the cluster.
    \item The projected area of the Faraday depth enhancement extends 2--4 times the projected distance of the X-ray emitting ICM, though is mostly contained within the cluster's virial radius.
    \item The global Faraday depth enhancement naturally divides into two distinct (projected) morphological sub-regions: (1) A triangular region extending from its vertex near NGC 1399 towards its apparent base about $\sim1.5^\circ$ to the NE, with a mix of both positive and negative $\phi_{\mathrm{peak,res}}$ values, and (2) a banana-shaped strip of width $\sim0.5^\circ$ and length $\sim2.5^\circ$, curving slightly around NGC 1399, but centred $\sim0.75^\circ$ to its SW, and having predominantly negative $\phi_{\mathrm{peak,res}}$ values. 
    \item On average, the areal total and polarised radio emission density is $\sim30\%$ and $\sim50\%$ lower within 1 degree (360 kpc) of the Fornax cluster (respectively) compared to outside this radius. Cumulative source counts versus flux density show that the emission decrement persists over the full range of flux densities that are effectively probed by our observations and sample. 
\end{itemize}

\section{Discussion}\label{sec:disco}

\subsection{Location of the Faraday screen along the line of sight}\label{sec:location}

There are two possible locations for the excess Faraday rotating material identified in Section \ref{sec:rms}: In our Galaxy, or in the Fornax cluster. The field contains intervening Galactic emission in {\sc H\,i}, H$\alpha$ and soft X-rays. The interstellar medium thereby traced is Faraday-active, appearing capable of inducing mild depolarisation via \emph{differential} Faraday rotation in spatially-limited regions due to its arcsecond-scale ionisation structure \citep{Anderson2015}. The question is whether degree-scale Galactic foregrounds can enhance variance in \emph{net} Faraday rotation over arcminute scales. We consider this unlikely, because:

\begin{itemize}
    \item \citet{Anderson2015} showed that in this field (i.e. centred on Galactic $l,b$ of 235.988$^\circ$, $-$55.484$^\circ$), over spatial scales of several arcminutes, the expected contribution of the Galactic foreground to RM variance is only $\sim50$ rad$^2$m$^{-4}$ (see Figure 36 and Section 7.2.1 of that work). The $282^{+86}_{-75}$ rad$^2$m$^{-4}$ variance in the vicinity of the cluster is $\sim6\pm2$ times higher than this.
    \item We cannot identify any foreground structures in maps of Parkes Galactic All Sky Survey \citep[GASS;][]{McClure-Griffiths2009} {\sc H\,i} column density, \emph{Planck} spectral brightness, \emph{ROSAT} soft X-rays, or H$\alpha$ photon flux that map to precisely the same region in which the Faraday depth enhancements occur (though structure in such emission is present in the field --- see Figure 37 of \citealp{Anderson2015}).
    \item The prior probability that a degree-scale Galactic Faraday depth enhancement would happen to centre precisely on a pre-existing and well-defined position of interest is not well known, but must surely be small.
\end{itemize}

\noindent In what follows then, we assume that the Faraday active material is physically associated with the Fornax cluster.

\subsection{Mass and phase of the ionised cluster gas}\label{sec:mass}

Little is concretely known about the structure of magnetised plasma in the periphery of low-mass galaxy clusters, including the strength and characteristic scale-length of the turbulent and regular magnetic fields in the rarefied plasma. In the case of the Fornax cluster, our difficulties are compounded by the fact that the Fornax cluster is not relaxed, but is being disturbed by ongoing mass-assembly processes. This can distribute $\mu$G-level magnetic fields throughout cluster volumes (e.g. \citealp{Xu2009}), inducing significant variability in the Faraday rotation signal of clusters with otherwise similar properties \citep{Xu2011}, and degeneracies in the interpretation of such data \citep{Johnson2020}. Nevertheless, following \citet{Anderson2018}, we can crudely estimate the baryonic gas mass generating the Faraday depth enhancement as follows. The peak Faraday depth of a column of thermal electrons threaded by a uniform magnetic field is (e.g. \citealt{HH2012}):

\begin{eqnarray}
\phi &=& 0.81n_eB_{u,||}L \nonumber \\
&=& 26N_{e,20}B_{u,||} ~\text{rad m}^{-2}
\label{eq:patchFD}  
\end{eqnarray}

\noindent where $B_{u,||}$ is the strength of the magnetic field projected along the line of sight [$\mu$G], and $N_{e,20}$ is the electron column density in units of $10^{20}$ cm$^{-2}$. From Figure \ref{fig:fornax_voronoi}, we estimate the solid angle occupied by the Faraday depth enhancements to be $4.1\times10^7$ square arcseconds, corresponding to a projected area $A=3.8\times10^{48}$ cm$^{2}$ at the distance of the cluster. Assuming the Faraday-active plasma is dominated by ionised hydrogen and helium nuclei with a typical ICM abundance ratio of 9:1 (respectively), and thus a mass density of $1.17n_em_{p}$ (where $m_{p}$ is the mass of a proton), the total mass of baryonic material is roughly:

\begin{eqnarray}
M_\text{Thermal}&\approx&1.17\times10^{20}N_{e,20}Am_\text{p}
\label{eq:patchFD2}  
\end{eqnarray}

\noindent Defining the characteristic Faraday depth of the material as $\phi_{\mathrm{peak,res,char}}$, then combining Eqns. \ref{eq:patchFD} and \ref{eq:patchFD2} yields:

\begin{eqnarray}
M_\text{Thermal}&\approx&1.39\times10^{10}\bigg(\frac{\phi_{\mathrm{peak,res,char}}}{B_{u,||}}\bigg) ~M_\odot
\label{eq:patchFD3}  
\end{eqnarray}

\noindent Using  $\phi_{\mathrm{peak,res,char}}\sim\sigma_{\phi_{\mathrm{peak,res,cluster}}}\approx17$ rad m$^{-2}$, we get $M_\text{Thermal}\approx2.4\times10^{11}\big(\frac{1}{B \mu\text{G}}\big)~M_\odot$, which for comparison, is approximately three times the mass of the X-ray-emitting hot ICM (but see \citealp{Johnson2020} for discussion of inherent uncertainties in such measurements). If there are typically $N$ magnetic field reversals along the line of sight to sources in our sample, $M_\text{Thermal}$ will be larger by a factor of $\sim N^{1/2}$.

The material extends well beyond the presently-detectable hot ICM, typically to a distance of 360 kpc. This yields an average thermal electron density inside this volume of $n_e\approx4.3\times10^{-5}\big(\frac{1}{B \mu\text{G}}\big)N^{1/2}f^{-1}$ cm$^{-3}$, where $f$ is the volume-filling factor. If $B$, $N$, and $f$ are equal to 1, it represents a baryonic over-density of $\delta\sim215$ --- a regime which is obviously very different to the hot ICM, for which $\delta\gg1000$ for the inner regions of the cluster, but which drops off precipitously to $\delta\sim50$ by 180 kpc from the cluster centre, and which is projected to drop by a further two orders of magnitude to negligible levels by 360 kpc (based on \emph{ROSAT} measurements of soft X-ray emission; see Sections 2.6 and 2.6 of \citealt{Paolillo2002}). Instead, it is comparable to the radially-averaged characteristic density of the intergalactic medium expected to inhabit $\sim10^{13}~M_\odot$ galaxy groups and poor clusters \citep{Dave2010,Haan2014}, and the moderately-dense phases of the WHIM thought to surround galaxy clusters \citep{Dave2001,Yoon2012}. Alternatively, it is possible (or perhaps likely) that the magnetic field strength $B$ is lower than 1 $\mu$G, or that the number of magnetic field reversals along typical lines of sight $N$ is large, or that the volume filling factor $f$ is small, or some combination of the above. But it each case, this would tend to raise our total gas mass estimate. While this might bring the baryonic over-density more into line with expectations for a hot ICM phase in a cluster environment, its radial distribution would be rendered even more inconsistent with direct measurements of radial gas densities in this hot phase reported by \citet{Paolillo2002}. Thus, we claim to have directly detected a moderately-dense phase of the diffuse WHIM via Faraday rotation, which is either too cool, too diffuse, or both, to have been detected by X-ray observatories to date (though this may be addressed with future observations by eROSITA for example; \citealp{Merloni2012}). This appears to be made possible by cluster dynamic processes that act to organize and amplify the embedded magnetic field. We discuss this further in Section \ref{sec:nature}. 

\subsection{Structure and origin of the Faraday depth enhancements}\label{sec:nature}

The morphology of the Faraday depth enhancement (Section \ref{sec:spatial}; Figures \ref{fig:scenario},\ref{fig:scenario_drawing}) is reminiscent of the bow shocks seen around merging cluster cores (e.g. \citealp{MV2007}), and astrophysical bow shocks more generally. If a shock system does exist, it could be either a stationary bow-shock caused by interaction between the NE and SW sub-clusters \citep{Drinkwater2001,Scharf2005}, or a propagating shock, set up by merger activity in the main (i.e. NE) sub-cluster itself \citep{Sheardown2018}. We consider these possibilities in turn.

\subsubsection{NE-SW sub-cluster merger}\label{sec:subcluster_merger}

\citet{Drinkwater2001} proposed that the Fornax cluster can be partitioned into NE and SW clumps, which are likely merging at speeds between 100 and 500 km s$^{-1}$. \citet{Scharf2005} interpreted the swept-back (to the NE) morphology of the X-ray-emitting main cluster ICM as evidence to support this. Extrapolating X-ray-derived temperature and density measurements from \citet{Jones1997} to the approximate radius of the shock front (1 degree) yields an estimated sound speed of $c_s=1480\sqrt{T/10^8~\text{K}}\approx420$ km s$^{-1}$ \citep{Sarazin2002}, and an Alfv\'en speed of $c_A=\sqrt{B^2/\rho}\approx320$ km s$^{-1}$. The merger speed $V$ may therefore be transonic, in which case a bow shock could form in diffuse gas between the merging components. 

\begin{figure*}
\centering
\includegraphics[width=0.75\textwidth]{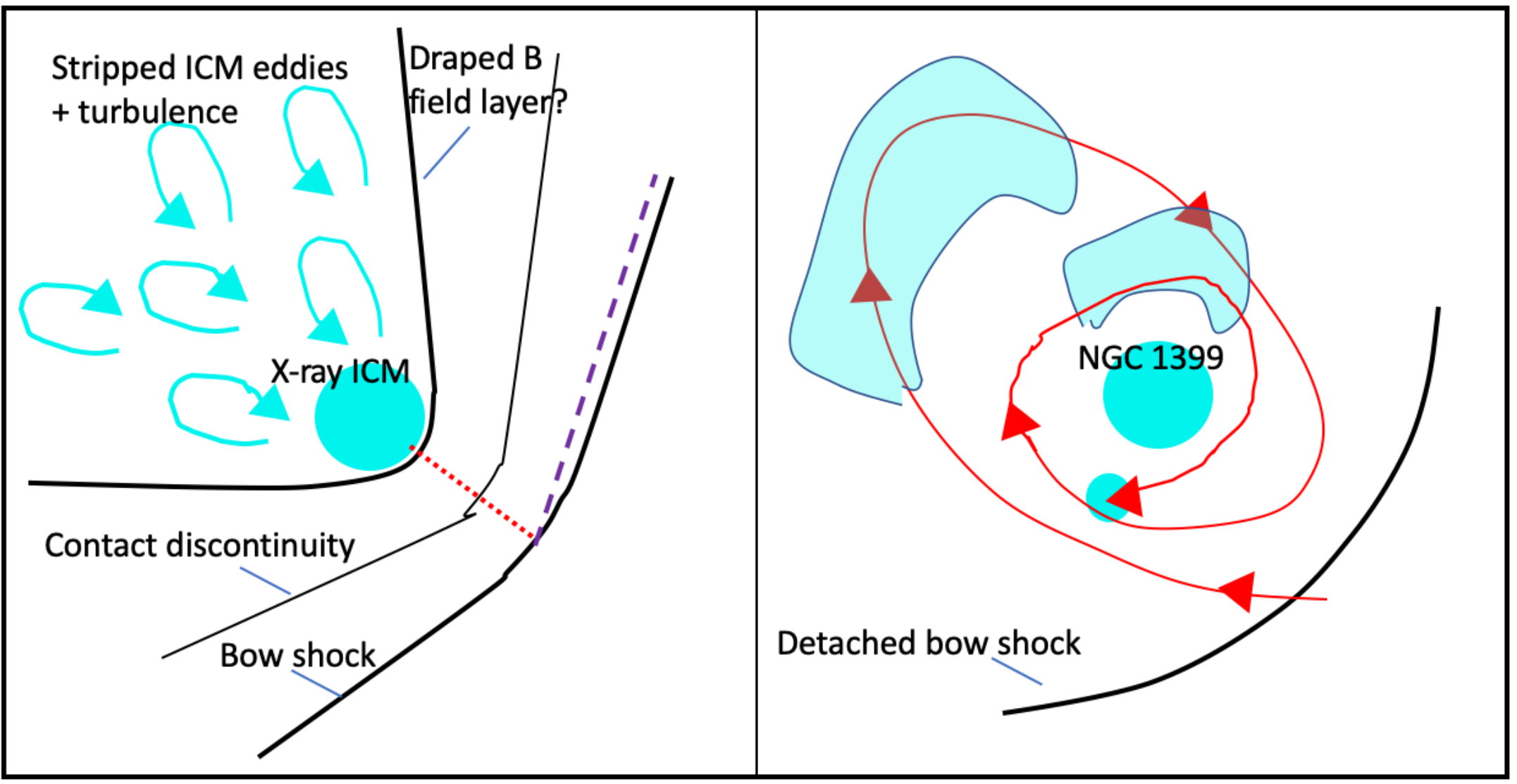}
\caption{Physical features for the NE-SW sub-cluster merger (left panel) and NGC 1404 merger (right panel) scenarios, described in Section \ref{sec:nature}. \emph{Left panel:} Light blue filled circle: X-ray emitting ICM; Light blue arrows: Turbulent eddies in stripped ICM; Black lines: Canonical features of astrophysical shocks, as discussed in the main text; X-ray emitting ICM; Red dotted line: Shock stand-off distance $d_s$; Purple dashed line: Projected angle of shock front. Note that the angle between the red-dotted and purple-dashed lines is the Mach angle referred to in Section \ref{sec:subcluster_merger}. \emph{Right panel:} Large light blue filled circle: NGC 1399; Small light blue filled circle: NGC 1404; Light blue blobs: Wake features generated by NGC 1404 in-fall, as described in the main text; Red line and arrows: Path of in-falling NGC 1404; Black line: Detached moving bow shock generated by NGC 1404 in-fall.}
\label{fig:scenario_drawing}
\end{figure*}

We propose that the features apparent in our maps of $\phi_{\mathrm{peak,res}}^2$ (Figure \ref{fig:scenario}) may correspond to the canonical features of a stationary shock leading a blunt object as illustrated in Figure \ref{fig:scenario_drawing} (left panel). That is, in this instance (a) a concave shock front centred around the X-ray-emitting ICM, whose point of closest approach is located 360 kpc to its SW, (b) a contact discontinuity between the main cluster gas and that in a more diffuse extended envelope, located perhaps one third of the way from NGC 1399 to the leading shock, and (c) a triangular extension of the X-ray-emitting ICM to the NE, representing a less-dense and less-hot phase of the magneto-ionised ICM stripped from the NE cluster. The boundary of this region may be associated with magnetic fields draped over the X-ray emitting ICM and amplified there --- expected to appear for any super-Alfv\'enic merger in magnetised media \citep{Lyutikov2006}. 

If the features described in the previous paragraph are correctly identified, then since the radius of the X-ray-emitting medium is small compared with the length of the shock front, we can estimate the Mach number of the shock via its Mach angle and stand-off distance (see Figure \ref{fig:scenario_drawing}). The Mach angle is given by $\mu=\text{sin}^{-1}({1/M_s})$, where $M_s$ is the sonic Mach number. From visual inspection of Figure \ref{fig:scenario}, $\mu\approx70\pm5^\circ$ yielding $M_s=1.06^{+0.04}_{-0.03}$, placing the merger in the transonic regime. The shock stand-off distance $d_s$ is a sensitive and independent measure and test of this claim \citep{Sarazin2002} --- it depends only on the value of $M_s$ and the effective shape of the supersonic object. A prediction for the stand-off distance versus Mach number is given by Sarazin (2002, Figure 4). For our best estimate of $M_s=1.06$, we expect $d_s/R\approx7$, where $R$ is the radius of curvature of the dense core. For the dense X-ray-emitting ICM, we estimate $R\approx7$ arcminutes in the SW direction using X-ray contours provided by \citet{Paolillo2002} and \citet{Scharf2005} (also see Figure 5), and so expect the bow shock front to be located $\sim49$ arcminutes SW of the cluster. This is essentially consistent with the location of the proposed shock front in our maps (1 degree to the SW of the cluster core). 

This sub-cluster merger model predicts several features that could be used to confirm or refute it in future, and which might then be used to measure the properties of the gas. First, a generic feature of super-Alfv\'enic mergers is that magnetic fields should be `collected' and amplified along a thin conic layer by the dense cluster ICM upon its passage through the more extended diffuse envelope. This is known as magnetic draping \citep{Lyutikov2006}, and is expected to occur for a wide range of merger speeds and magnetic field configurations. The expected minimum thickness for the magnetic field amplification layer is $\sim2Rc_A/V\approx10$ arcminutes \citep{Lyutikov2006}, presumably draped over the leading edge of the dense core $\sim7$ arcminutes to the SW of NGC 1399. At such a pressure-balance surface, we expect that (a) a spatial discontinuity in X-ray emission may be observed (e.g. \citealp{Machacek2005}), and (b) the magnetic pressure will roughly equal the ram pressure in the zone of plasma compression, such that $B^2\sim8\pi\rho V^2$. For the NE-SW merger speed and ICM densities cited above, this implies a magnetic field strength of $\sim8~\mu$G. Both expectations are interesting, for they have been raised by \citet{Su2017} in the context of a sloshing cold front identified in this system at approximately the expected location (identified as the "F2" front in that work), and the hypothesis that such a cold front can be stabilised against Kelvin-Helmholtz instabilities via magnetic tension, if only strong enough fields can be generated. The stabilisation requires a magnetic field running locally parallel to the front with strength $\sim10\mu$G \citep{Vikhlinin2001}, which is evidently similar to the value we obtain for ram-pressure balance in the system. Thus, in our merger model, we suggest that the cold front identified by \citet{Su2017} may act as the effective surface of ram-pressure balance with a diffuse ionised medium through which the dense core of the cluster moves with transonic speed. The cold front might stabilize itself via the very magnetic draping and compression that the NE-SW sub-cluster merger induces.      

A second feature that might be sought to furnish evidence for or against our model is as follows. In the wake of blunt objects embedded in a hydrodynamic flow where the Reynolds number exceeds $\sim50$ (most likely applicable throughout galaxy clusters --- e.g. \citealp{Zhuravleva2019}), vortex shedding occurs (e.g. \citealp{lienhard1966}). This terms refers to a phenomenon where vortices are created behind a blunt object embedded in a fluid flow, and are periodically jettisoned into the flow. The vortex shedding frequency ($f$) is related to $L$ and the flow speed of the external fluid ($U$) by the Strouhal number ($St$) as $St=(fL)/U$, where $St\approx0.2$ over a wide range of Reynolds numbers (e.g. \citealp{Sakamoto1990,Gruszecki2010}). This defines a characteristic longitudinal scale for the vortex separation, while the transverse scale of the vortices is comparable to that of the blunt object ($L$). Thus, in the wake of the dense Fornax cluster ICM, we expect to see coherent vortical structures ranging in size from $\sim14$--70 arcminutes (84--420 kpc). This may also correspond to an injection scale for a subsequent turbulent cascade, whose detailed character and volume filling factor will depend on the effective Reynolds number of the medium (among other factors) (e.g. \citealp{Subramanian2006,Zhuravleva2019}). These observational signatures are not detectable in our present observations, but may be in pending deep observations of the region (see Section \ref{sec:concluding}).

Finally, we note that in Section \ref{sec:spatial}, we proposed that an enhancement in Faraday depth can be seen running adjacent to the NE of the long axis of the Fornax A lobes. \cite{Ekers1983} previously argued that the Fornax A system was moving to the N or NE due to a $\sim$ southerly offset of a diffuse radio bridge between the lobes and its radio core. Both this feature, and the purported Faraday depth enhancement to the NE, might be naturally explained as a consequence of gas and magnetic field compression in the sub-cluster merger model.

\subsubsection{NGC 1404 in-fall}\label{sec:ngc1404_infall}

Recently, \citet{Sheardown2018} simulated the in-fall and merger of the galaxy NGC 1404 to the main Fornax cluster, and highlighted several features that bear an intriguing correspondence with features in our data. Their preferred simulated in-fall scenario produces: (a) a detached bow shock from NGC 1404's previous orbit in the cluster potential, having a Mach number in the range 1.3--1.5, currently propagating outwards between 450 and 750 kpc S and SW of NGC 1399, and (b) regions of relatively cold turbulent material distributed through much of the cluster, but particularly towards the NE, again stirred up by NGC 1404's previous orbits through the system (see Figure \ref{fig:scenario_drawing}, right panel).

Qualitatively, these predictions are similar to the sub-cluster merger scenario above, and can explain our results almost equally well. A minor point of disagreement is that our putative shock is located 20\% closer to NGC 1399 than their lower limit on this distance. Nevertheless, we are reluctant to claim that our results strongly favour either one of the models with the present data. In the near future, the MeerKAT Fornax Survey \citep{Serra2016} will obtain much deeper observations, and may be able to trace (via Faraday rotation or depolarisation) the trails of cold gas and turbulence that \citet{Sheardown2018} predict will follow in the wake behind NGC 1404.

\subsection{Comparison with cluster stacking experiments}\label{sec:comparo}

The excess scatter ($\sigma_{\phi_{\mathrm{peak,res,cluster}}}=16.8$ rad m$^{-2}$) and spatial extent (360 kpc radius) of the RM enhancement are both smaller than reported for ensembles of galaxy clusters in RM stacking experiments. \citet{Clarke2001} report a standard deviation in RM of 114 rad m$^{-2}$ for sources located within the projected radius of galaxy clusters, compared to 15 rad m$^{-2}$ for a control sample, which we note is similar to our value of 11.8 rad m$^{-2}$ for our field sources. More recently, \citet{Bohringer2016} report a value of 120 rad m$^{-2}$ for sources located within the projected radius of clusters. However, \citet{Bohringer2016} also show that this scatter is driven mainly by massive clusters: When their sample is split by the median X-ray luminosity of their associated cluster (corresponding to a total mass of $1.4\times10^{14}$ M$_\odot$), the standard deviation in RM increases to $158\pm34$ rad m$^{-2}$ for sources associated with clusters in the upper half of the mass range, and decreases to $62\pm11$ rad m$^{-2}$ for the converse sources. Our lower derived value of of $16.8\pm2.4$ rad m$^{-2}$ for the RM scatter seems broadly consistent with this coarse mass dependence, given that the total mass of the Fornax cluster is only $6^{+3}_{-1}\times10^{13}$ M$_\odot$ \citep{Drinkwater2001,Nasonova2011,Maddox2019} inside a few Mpc, which is half the mass cut value used by \citet{Bohringer2016}. It is also comparable to results from simulations for unrelaxed clusters of approximately the same mass (see Figure 14 of \citealp{Xu2011}, for example). However, in light of our conclusions about the nature of the ionised gas from Section \ref{sec:mass}, B{\"o}hringer \emph{et. al}'s statement that the relationship between RM scatter and cluster mass confirms {\em "that the observed excess scatter in the RM in the lines of sight of galaxy clusters is due to the cluster ICM"} may require a minor qualification, in the sense that some portion of the RM scatter may be driven by plasma regimes that differ from the canonical hot ICM.

The \citet{Clarke2001} and \citet{Bohringer2016} studies, and results from other recent works of simulation and observation (\citealp{Marinacci2018} and references therein), show that cluster-based RM enhancements are typically a strong function of radius. The central 200 kpc of the cluster ensemble shows |RM| values often exceeding 100 rad m$^{-2}$, but these values decrease roughly exponentially until they become indistinguishable from field source RMs beyond $\sim800$ kpc. The behaviour of the Fornax cluster RMs appear to be qualitatively different: The degree of RM scatter is quite uniform out to a projected radius of 360 kpc, at which point it drops precipitously to roughly half its value (Figures \ref{fig:peak_FD_vs_impact_param} and \ref{fig:fornax_cluster_sliding-median_abs_RM}), and thereafter remains constant with projected radius. This qualitative difference could be due to several factors, possibly including that (a) the plasmas traced by our RMs compared to the stacking experiments differ in their phase or properties at various fiducial cluster-centric radii, most plausibly because the Fornax cluster is exceptionally poor in terms of the number of member galaxies, gas mass, and total mass (b) the Fornax ICM is being disturbed by ongoing mass assembly, or (c) clusters that contribute to stacking experiments may individually show similar behaviour, but this signal is averaged out in the stacking. Further observations of a range of individual clusters --- now possible with ASKAP and other instruments --- are obviously the key to exploring these possibilities. 

\subsection{Comparison with studies of individual low-mass galaxy clusters}\label{sec:comparo_indiv}

What studies do exist of the magnetised ICM in individual low-mass clusters generally involve mapping the RMs across the lobes of the central dominant radio galaxy (e.g. \citealp{Perley1984,Laing2008,Guidetti2010,Govoni2017}). This provides an exquisite view of magnetic structure in the inner-most regions of the ICM, but little information beyond the range where the the radio lobes are bright enough to calculate reliable RMs (10s--100 kpc). In addition, the fields are likely modified by interactions with the radio galaxies themselves \citep{Guidetti2012}.
RM grid studies avoid both of these problems, since the sources are not in physical proximity to the medium being probed, and generally provide abundant information beyond 100 kpc, but limited information inside this distance. While there are not yet sufficient data to connect the results of these experiments across the gap in scales, some brief comments are nevertheless warranted.

\citet{Laing2008,Guidetti2010} and \citet{Govoni2017} mapped the RMs of centrally-located radio lobes to study the magnetised ICM of poor clusters with $\sim$ hundreds of member galaxies, masses in the range 4--$9\times10^{13}~M_\odot$ \citep{KB1999,Nikogossyan1999}, and central magnetic fields strengths of 3--10 $\muup$G, making them broadly comparable to the Fornax cluster. These studies measure RM dispersions of 0--40 rad m$^{-2}$, which our measured enhancement of 17 rad m$^{-2}$ falls squarely inside. For those studies however, the RM dispersion measured against the lobes generally drops to a few rad m$^{-2}$ by cluster-centric distances of 100 kpc (see Figures 9 and 12 of \citealp{Laing2008} and \citealp{Guidetti2010} respectively, and Figure 14 of \citealp{Govoni2010}), whereas the Faraday depth enhancement in the Fornax cluster is maintained out to at least 360 kpc. The studies cited above also attributed the enhanced Faraday dispersion to the cluster ICM, and indeed, the scale-size of X-ray emission from the hot ICM and region of enhanced Faraday depth are well matched, unlike the case for the Fornax cluster. It is not clear how to reconcile these results with the present data. It is possible that the Fornax cluster is being fed with more warm-hot gas from filaments of large-scale structure, or that similar reservoirs of gas exist in the vicinity of the other clusters but have not been detected yet, or that this gas exists but can only be detected in certain situations --- in the case of the Fornax cluster we speculate, because its ongoing merger status organises the magnetic field structure to as to induce a detectable Faraday rotation signature over a sufficiently large area of sky. Further observations of a larger sample of similar clusters are clearly needed.

\citet{Laing2008,Guidetti2010} speculate that the outer scale of magnetic field fluctuations --- of order 70 kpc in both cases --- could be set by the characteristic separation of interacting cluster members. If this picture is correct, then the $\sim3$ degree separation of the merging NE and SW Fornax sub-clusters might have been used to predict the $\sim$degree-scale Faraday depth enhancements that we have detected in the cluster \emph{a priori}. In turn, this may suggest that clusters undergoing similar mergers of massive sub-components, such as the nearby Antlia (e.g. \citealp{CR2015}, and references therein) and Centaurus \citep{Churazov1999} clusters, may represent particularly valuable targets for future RM grid studies, since the detectable x-ray-emitting ICM may not provide the whole picture. Finally, in Section \ref{sec:subcluster_merger}, we predicted that if the Faraday depth enhancement to the NE of the cluster is associated with vortex shedding of the cluster gas, we might expect to see coherent associated RM structures ranging in size from $\sim14$--70 arcminutes (84--420 kpc). This scale-size range is apparently characteristically larger than the RM structures induced by quiescent evolution in individual poor, relaxed, low-mass clusters, and so if RM structure is observed on these scales in the Fornax and other clusters, it might act as a signpost of ongoing merger and gas-stripping activity.

\subsection{The RM grid as a sensitive tracer of `missing' baryons?}\label{sec:bary}

Given our conclusions that the Faraday-active ionised gas is associated with the WHIM, it is interesting to consider whether Faraday rotation measure grids (e.g. \citealp{Gaensler2004}) and associated advanced analysis techniques (e.g. \citealp{Vernstrom2019,OSullivan2020,Stuardi2020}) might effectively probe such material. The thermal electron density drops rapidly with cluster-centric radius in typical ICM models, and the efficiency of Bremsstrahlung as a tracer of this gas drops more precipitously, since its emissivity is proportional to $n_e^2$. In contrast, the Faraday effect is proportional to $n_eBL$, assuming no significant separation of magnetic fields and thermal electrons along the line of sight. It is not clear what structure magnetic fields generally have at the periphery of clusters, but it is clear that certain processes can amplify magnetic fields, increase their degree of order, push their auto-correlation lengths to larger characteristic scales, or simply inject large-scale magnetic fields to begin with (e.g. sub-cluster mergers; e.g. \citealp{Govoni2005,Bonafede2009,Girardi2016}, magnetic draping; e.g. \citealp{Lyutikov2006}, shocks; e.g. \citealp{vanWeeren2010,Bruggen2011}, subsonic gas motions and shear flows; e.g. \citealp{Keshet2010,ZR2016,Donnert2018}, also this work). Finally, the periphery of clusters provides Mpc-scale path lengths through which polarised emission can propagate through such plasmas. Therefore, it seems reasonable to conclude that Faraday rotation measures should provide a sensitive means of detecting and studying rarefied warm plasma phases, assuming a sufficient density of background sources over a wide enough area. The current generation of radio telescopes will soon provide this routinely for many galaxy clusters.

\subsection{Evaluation of the decrements and enhancements in source counts and flux observed near the Fornax cluster}\label{sec:discoreduced}

Following the results and arguments presented in Section \ref{sec:linpol}, there are two two possible causes for the polarised and total intensity emission decrements observed inside one degree cluster-centric radius: (1) scattering or absorption by foreground material, or (2) a genuine paucity of radio emission due to cosmic variance. 

The former possibility requires that cool, dense gas lies along the line of sight. While it can be shown that the required gas could plausibly exist in the cluster, perhaps in the form of cloudlets of $\sim10^4$ K photoionised gas that often envelop elliptical galaxies (e.g. \citealp{Gauthier2010,Thom2012,Prochaska2013,Farnes2017,Lan2017,Berg2019,Pradeep2019,Werner2019,LR2020}), we can test this scenario using the frequency-dependence of the effect. Consider that the free-free optical depth is given by

\begin{equation}
\tau_{\text{ff}}\approx3.27\times10^{-7}\bigg(\frac{T_e}{10^4~\text{K}}\bigg)^{-1.35}\bigg(\frac{\nu}{\text{GHz}}\bigg)^{-2.1}\bigg(\frac{\text{EM}}{\text{pc cm}^{-6}}\bigg)
\label{eqn:ffabs}
\end{equation}

\noindent where $T_e$ is the electron temperature, $\nu$ is the observing frequency, and EM is the emission measure. 

We therefore cross-matched the discrete sources from our sample found within 1 degree of the cluster centre with radio surveys over a wide range of frequencies (including the GLEAM survey (72--231 MHz; \citealp{Wayth2015,Hurley-Walker2017}) survey, the VLA Low-frequency Sky Survey redux (VLSSr; 74 MHz; \citealp{Lane2014}), the TIFR GMRT Sky Survey (TGSS; 150 MHz; \citealp{Intema2017}), the Westerbork in the Southern Hemisphere (WISH; 352 MHz;  \citealp{DeBreuck2002}) survey, the Molonglo Reference Catalogue (MRC; 408 MHz;  \citealp{Large1981}), the Sydney University Molonglo Sky Survey (SUMSS; 843 MHz;  \citealp{Bock1999,Mauch2003}), the NVSS (1.4 GHz; \citealp{Condon1998}), the Parkes-MIT-NRAO (PMN; 4.85 GHz;  \citealp{Gregory1994,Wright1994}) survey, the Australia Telescope Parkes-MIT-NRAO (ATPMN; 5 and 8 GHz; \citealp{McConnell2012}) follow-up survey, and the Australia Telescope 20 GHz (AT20G; \citealp{Murphy2010})) to reconstruct their total intensity spectra. The result is that no sources brighter than 10 mJy/beam at 887 MHz in this sub-sample possess a spectral peak at frequencies above $\sim100$ MHz, and we conclude that the available evidence does not support the absorption hypothesis.

Considering the cosmic variance scenario in turn, randomly sampling $\pi$ square degrees of sky down to $\sim$mJy beam$^{-1}$ sensitivity (at which we continue to see the emission decrement of up to $\sim50\%$; see Figure~\ref{fig:source_counts}) yields a $1\sigma$ uncertainty in source counts of only $\sim15\%$ from cosmic variance (\citealp{Heywood2013}, Figure 2), corresponding to only a $\sim0.1\%$ probability that our results can be ascribed to cosmic variance alone. Nevertheless, decrements of similar depth and angular size have been observed in NVSS which, it has been claimed, may be associated with supervoids in the large-scale structure of the Universe (e.g. \citealp{Rudnick2007}). 

The enhancement in polarised and total radio flux, which appears to arc around the cluster at projected radii between 1 and $\sim2$ degrees, appears even more difficult to explain. The handful of redshift cross-matches in this region indicate distances much greater than the cluster, making it appear to be a chance effect. We are unaware of known clusters in these locations, but redshift coverage here is currently relatively sparse. 

Nevertheless, remarkable large-scale cosmic structure does appear to exist in both the foreground and background of the Fornax cluster: Purportedly, a dense filament of dark matter extends along the entire line of sight between the Fornax cluster and the Milky Way Galaxy (see Figure 4 of \citealp{Hong2020}), while the $\sim190\times90\times140$ Mpc diameter Sculptor void lurks immediately behind \citep{Tully2019}. At a near-side distance of only $\sim30$ Mpc, the latter possesses an apparent angular diameter which is far too large to explain our source decrement/enhancement structures, which cover less than 10 square degrees. Nevertheless, such voids appear to contain significant substructure, and are linked to other voids in a sponge-like lattice which cannot yet be mapped out in detail to the typical distance of powerful radio sources. We speculate that a lattice of connected under-dense regions could result in smaller-scale `tunnels', through which the average density and brightness of radio sources is lower on average, and perhaps chance alignments of higher-density filaments or walls where the opposite is true. With the available evidence failing to support scenarios involving instrumental effects, depolarisation, or absorption, we tentatively conclude that cosmic variance due to such large scale structure may be the most viable explanation for our results. Over the next several years, the Evolutionary Map of the Universe (EMU; \citealp{Norris2011}) and POSSUM surveys will map the radio sky with an unprecedented combination of sky coverage, depth, and detail. This will shed much new light on the 3-dimensional distribution of radio sources and large-scale cosmic structure, as well as on Faraday rotation associated with the gas in this cosmic web. Our result can then be revisited and reinterpreted if necessary.

\section{Conclusion}\label{sec:concluding}

We have conducted the first Faraday rotation measure grid study of an individual low mass galaxy cluster, achieving a polarised source density of 27 per square degree using the revolutionary new survey capabilities of the ASKAP telescope. Our key results are that:

\begin{itemize}
    \item The distribution of peak Faraday depths for confirmed background radio sources shows an excess dispersion of $\sim17$ rad m$^{-2}$ within 1 degree (360 kpc) of the Fornax cluster centre, and in more spatially-limited regions out to the 705 kpc virial radius of the cluster. This is between 2 and 4 times farther than the projected distance of the currently observable X-ray emitting ICM.
    \item We estimate that the mass of the Faraday-active gas is $2.4\times10^{11}~M_\odot$, which is approximately triple the mass of the hot ICM so far detected in X-rays, but with a low average density of $n_e\approx2.7\times10^{-5}$ cm$^{-3}$. This represents a baryonic over-density (compared to the cosmic average) of $\delta\sim215$ found at cluster-centric radii out to 360 kpc. This appears to be markedly different from the hot ICM phase in Fornax, for which $\delta\gg1000$ in the core of the cluster, but which is observed to drop to $\delta\sim50$ by 180 kpc, and projected to drop to negligible densities much beyond this.
    \item Morphologically, the Faraday depth enhancement divides into two regions. We interpret one as an astrophysical shock front, and the other as the turbulent main-cluster wake, in a scenario in which NE and SW sub-clusters are merging at transonic speeds. Alternatively, it may be that the protracted merger of the dominant cluster galaxy NGC 1399 and the in-falling galaxy NGC 1404 has produced a detached shock moving to the SW, and has redistributed cold turbulent gas from the cluster centre and the ISM of NGC 1404 throughout the cluster. 
    \item On average, the total and polarised radio emission areal density is $\sim30\%$ and $\sim50\%$ lower (respectively) within 1 degree of the Fornax cluster compared to outside this radius. Consistent with this, cumulative source counts versus flux density show the $\sim50\%$ emission decrement persists over a large range of flux densities effectively probed by our observations and sample. Depolarisation, instrumental effects and image artifacts, and free-free absorption by a cold and dense gas, are all ruled out as possible causes or found to be unlikely. Cosmic variance also appears to be an unlikely explanation, but one which we tentatively favour in lieu of more compelling evidence. This result should be examined using future, deeper observations.
\end{itemize}

We argued that the Faraday-active gas is associated with a moderately dense phase of the WHIM, in which ongoing merger processes in the cluster continue to amplify and organize magnetic fields, thereby providing ideal conditions to trace this material with the Faraday effect. In particular, a shock system traced by Faraday RMs may confirm that the NE and SW Fornax appear to be undergoing a transonic merger, as previously described. 

These results demonstrate that deep, wide-field RM grid studies have the capacity to reveal the gas in and around galaxy clusters in a diverse set of regimes. This being the case, impending deep all-sky linear polarisation surveys like POSSUM, the VLA Sky Survey (VLASS; \citealp{Lacy2019}), the POlarised GLEAM survey (POGS; \citealp{Riseley2018,Riseley2020}), and the LOFAR Two-Metre Sky Survey (LoTSS; \citealp{Shimwell2017}) will all help to revolutionize our understanding of large samples of such objects, while much deeper targeted observations like the MeerKAT Fornax Survey \citep{Serra2016} will reveal a wealth of structure in individual clusters that has gone heretofore unseen.   

Finally, in addition to the RM grid analysis presented here, \citet{Anderson2015} demonstrated the value of  \emph{depolarisation} grids to study small-scale magnetoionic structure along the line of sight. In this work, we initially postulated that ICM depolarisation may be responsible for the reduced counts and polarised flux we observed in the vicinity of the Fornax cluster, but then recognised that this signal was largely reflected in total intensity emission, too. It remains unclear where this underdensity comes from, but it is reasonable to assume that such fluctuations will be found on these angular scales elsewhere on the sky, for similarly uncertain reasons (e.g. see \citealp{Rudnick2007}). Now that deep polarised source grids are becoming available, which provide RM and depolarisation information at areal densities of $\sim$ several tens per square-degree, we must be aware of these fluctuations, and pursue a better understanding of the clustering scales for extragalactic radio sources.

\section*{Acknowledgements}

The authors would like to thank the anonymous reviewer for their time, and for their very helpful comments. Our paper has benefited significantly from this input. C. S. A. is a Jansky Fellow of the National Radio Astronomy Observatory. C. S. A. thanks Phil Edwards for helpful feedback which has improved this paper, and Joe Callingham for curating the multi-survey spectral measurements discussed in Section 4.2. The Australian SKA Pathfinder is part of the Australia Telescope National Facility which is managed by CSIRO. Operation of ASKAP is funded by the Australian Government with support from the National Collaborative Research Infrastructure Strategy. ASKAP uses the resources of the Pawsey Supercomputing centre. Establishment of ASKAP, the Murchison Radio-astronomy Observatory and the Pawsey Supercomputing Centre are initiatives of the Australian Government, with support from the Government of Western Australia and the Science and Industry Endowment Fund. We acknowledge the Wajarri Yamatji people as the traditional owners of the Observatory site. The POSSUM project has been made possible through funding from the Australian Research Council, the Natural Sciences and Engineering Research Council of Canada, the Canada Research Chairs Program, and the Canada Foundation for Innovation. The Dunlap Institute is funded through an endowment established by the David Dunlap family and the University of Toronto. B.M.G. acknowledges the support of the Natural Sciences and Engineering Research Council of Canada (NSERC) through grant RGPIN-2015-05948, and of the Canada Research Chairs program. CIRADA is funded by a grant from the Canada Foundation for Innovation 2017 Innovation Fund (Project 35999), as well as by the Provinces of Ontario, British Columbia, Alberta, Manitoba and Quebec. C. J. R. acknowledges financial support from the ERC Starting Grant ``DRANOEL'', number 714245. NMc-G acknowledges funding from the Australian Research Council via grant FT150100024. S.P.O. acknowledges financial support from the Deutsche Forschungsgemeinschaft (DFG) under grant BR2026/23. Partial support for LR is provided by U.S. National Science Foundation grant AST-1714205 to the University of Minnesota. J. M. S. acknowledges the support of the Natural Sciences and Engineering Research Council of Canada (NSERC) Discovery Grants program. 



\appendix

\section{Estimating the uncorrected off-axis polarisation leakage}\label{sec:appendixa}

The magnitude of uncorrected off-axis Stokes $I\rightarrow Q$ and $I\rightarrow U$ polarisation leakage can be estimated for ASKAP data using field sources themselves as probes, by adopting a modified version of the approach described by \citet{Lenc2017}, as follows.

Within the $\sim34$ square degrees covered by the ASKAP footprint, we detect 18640 sources in Stokes $I$ --- approximately 550 sources per square degree on average. We extracted the Stokes $I$, $Q$, and $U$ values at the location of each of each of these sources in multi-frequency synthesis mosaics of the field, and calculated the fractional Stokes quantities $q=Q/I$ and $u=U/I$. These data points provide a local point-probe of the effective frequency-independent leakage between the relevant Stokes parameters in our linear mosaics. We call them `local leakage estimators' here. 

We create maps of the Stokes $I\rightarrow Q$ and $I\rightarrow U$ polarisation leakage over the ASKAP footprint by first forming a dense, regular grid of coordinate locations covering the field. At each coordinate location, we then:

\begin{enumerate}
\item select all Stokes I sources with a full-band (i.e. 288 MHz) $\text{S/N}>10$ in an $0.4^\circ$ radius aperture around the point, corresponding to one quarter of the FWHM of the formed beams, and typically netting $\sim300$ source components
\item calculate the inverse-variance-weighted mean of Stokes [Q,U]/I for the sample, providing an initial estimate of the local polarisation leakage, which will be affected true polarised sources
\item calculate the standardized residuals (SRs) of Stokes [Q,U]/I from the initial leakage estimate for each source component --- that is, the number of standard deviations that each source's Stokes [Q,U]/I value lies away from the initial leakage estimate in both the positive and negative directions
\item identify sources whose SR falls in the upper and lower 5\% of the SR distribution, and remove these from subsequent calculations. These $\sim30$ source components are assumed to be genuinely polarised sources (i.e. their polarisation is inconsistent with the leakage model). The remaining ensemble of sources are assumed to be polarised only insomuch as they are affected by the local polarisation leakage, which will be roughly similar in manner and degree, and can therefore provide a reasonable estimate of such.
\item perform step (2) again with the trimmed sample, which results in our final estimate of the local polarisation leakage at the coordinate location in question
\end{enumerate}

\begin{figure}
\centering
\includegraphics[width=0.5\textwidth]{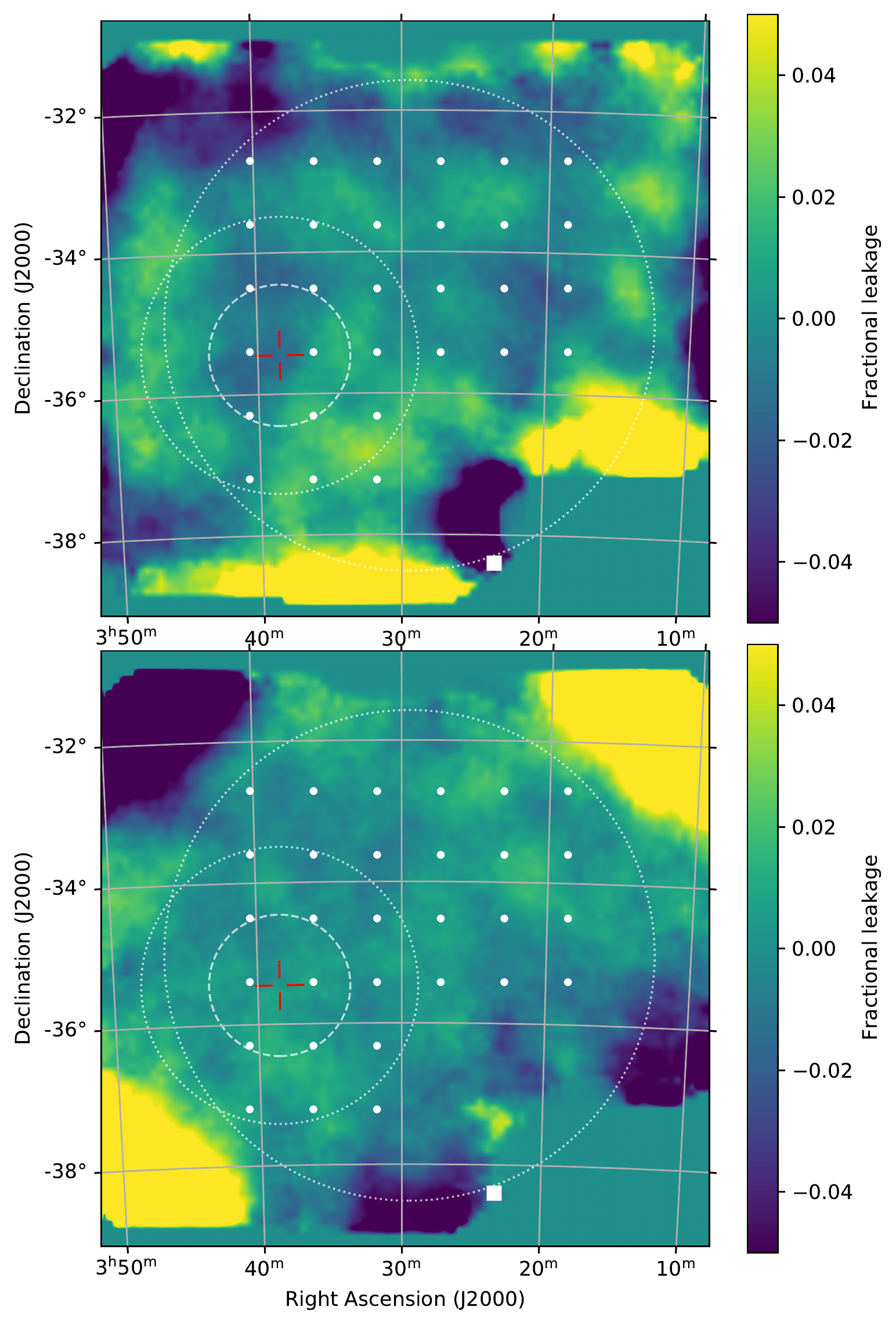}
\caption{Position-dependent Stokes $I\rightarrow Q$ (top) and $I\rightarrow U$ (bottom) leakage maps, derived as described in the main text. The centre of the Fornax cluster is indicated with a red cross-hair. The white dashed circle indicates an angular radius of one degree, the smaller of the two white dotted circles indicates the 705 kpc ($1.96^\circ$) virial radius of the cluster, and the larger of the two white dotted circles indicates $3.5^\circ$ distance from the centre of the mosaic. We do not consider sources in the main analysis of this paper, in order to reject regions of large $I\rightarrow U$ leakage from the corner beams. The white dots are the formed beam centres. For reference, the half-width-at-half-maximum of the formed beams is approximately equal to the separation of the beam centres (see Figure \ref{fig:peak_P_noise_beams}).}
\label{fig:leakage_higherrez_zoom_combined}
\end{figure}

\noindent The resulting maps for Stokes $I\rightarrow Q$ and $I\rightarrow U$ polarisation leakage over the full ASKAP beam footprint are shown in Figure \ref{fig:leakage_higherrez_zoom_combined}. Before highlighting the salient features of these maps, we first describe a set of simulations that we undertook to assess the accuracy of our estimates and methods. The simulator was set up to answer the following question: For a given grid location in our mosaic, and given (a) specified Stokes $I\rightarrow [Q,U]$ leakage value, (b) a randomly generated population of sources with realistic distributions of Stokes $I$, $Q$, and $U$ values and fractional polarisations, (c) the sensitivity of our observations, and (d) the leakage estimation protocol described above, what is the resulting distribution in the derived leakage values for 10,000 independent realisations of the simulation? The results were that our median estimated leakage value was almost identical to the specified input value ($0.98\times$ the specified input value, to be precise), while the standard deviation was 0.5\% --- that is, the uncertainty on the fractional leakage maps in Figure \ref{fig:leakage_higherrez_zoom_combined} is $\sim0.005$.

We now comment on the salient features of Figure \ref{fig:leakage_higherrez_zoom_combined}. As expected, the strongest leakage occurs at the mosaic edges. This is primarily due to the lack of adjacent beams contributing to mosaic beyond the outer ring, meaning that (a) the measurements reflect leakage values at much larger beam-centric radii than generally contribute to the mosaic within the outer ring of beams, and (b) the leakage experiences no averaging down in the linear combination of beams either. Quadrupolar symmetry is evident in both maps, which follows from the classic quadrupolar `clover leaf' leakage pattern displayed by the individual beam responses (offset by $45^\circ$ for Stokes q and u), which is characteristic of linear feeds. In the body of the mosaic, the leakage is typically less than 1\%, though values of up to 2\% are commonly observed in isolated regions, and can become as severe as 5\% for a handful of sources. We eliminated leakage-dominated sources from our sample as described in Section \ref{sec:obs}. 

\section{Details of the binning experiment from Section 5.2}\label{sec:appendixb}

\begin{figure}
\centering
\includegraphics[width=0.45\textwidth]{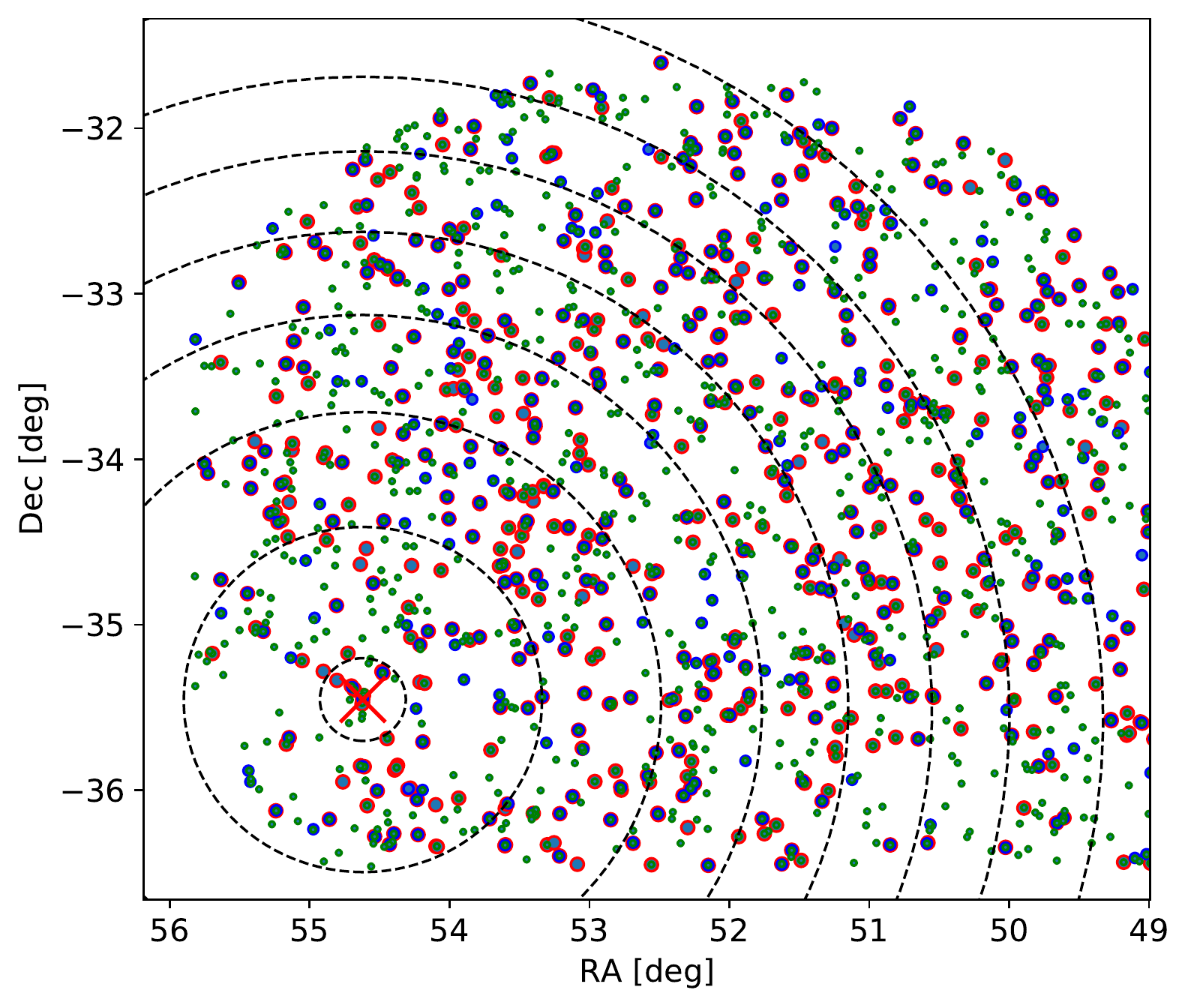}
\caption{Locations of the sources and bins used in the experiment described in Section \ref{sec:linpol} and Appendix B. Dots indicate the locations of polarised ASKAP (large red), polarised NVSS (medium blue), and GLEAM (small green) sources after the spatial truncations described in this appendix. The black dashed lines indicate the locations of the annular bins, each incorporating an effective area of $\pi$ square degrees (with the exception of the central and outer bins, which are not included in the analysis). The red `x' indicates the position of the Fornax cluster centre.}
\label{fig:ASKAP_NVSS_GLEAM_in_annular_bins}
\end{figure}

For the equal-area annular binning experiment described in Section \ref{sec:linpol}, the bin radii cannot be calculated analytically, because some of the annuli overlap truncated regions near the edge of our mosaic (see Sections 2 and 3). Therefore, to simplify the experiment, and to ensure that the annular bounding radii were as uniformly spaced as possible, we started by selecting an appropriate sub-set of our sample. Consider the following logical statements about the possible locations of sources in our mosaic. A source may be: (1) located outside a projected cluster-centric radius of 10 arcminutes, (2) located outside a projected cluster-centric radius of 10 arcminutes, but inside a cluster-centric radius of 1.014 degrees (corresponding to the first $\pi$ square degree annulus), (3) located outside a cluster-centric radius of 1.014 degrees, (4) located eastward of the western-most source satisfying condition 2 above, (5) located northward of the southern-most source satisfying condition 2 above, (6) located in the southeast quadrant relative to the cluster centre. We selected sources for this experiment satisfying the following logical combination of these conditions: (1 and 2) or ( (3 and 4 and 5) and not 6). The positions of the selected sources, and the NVSS and GLEAM sources also used in the experiment, are shown in Figure \ref{fig:ASKAP_NVSS_GLEAM_in_annular_bins}. Note that condition 1 ensures that we exclude known Fornax cluster sources from the analysis, and that these selection criteria were applied after, and in addition to, the selection criteria described in Sections 2 and 3.

Next, we calculated the bin radii corresponding to an effective (i.e. post-spatial-truncation) enclosed area of $\pi$ square degrees in each annulus. We did so by generating a dense grid of 10$^7$ points over the entire mosaic area, being random in location but uniform in their average spatial density. We then applied the same spatial truncations to these points as to the sample. By tallying up these random points, it is trivial to define the bin radii corresponding to the desired $\pi$ square degree effective bin areas. The resulting bounding radii for the annular bins are located 0.167, 1.014, 1.737, 2.324, 2.826, 3.313, 3.764 and 4.309 degrees from the cluster centre, and are shown in Figure \ref{fig:ASKAP_NVSS_GLEAM_in_annular_bins}.

The outcome of the experiments depend on the confidence intervals plotted in Figure \ref{fig:counts_plus_flux_in_annular_bins_mod3}. Note that our aim is to establish whether the \emph{observed} properties of sources behind the cluster differ significantly from those of the broader radio source population, based on our estimates of the mean and variance of the latter. We are \emph{not} trying to estimate the characteristics of the broader radio source population from the statistics in our individual $\pi$ square degree annuli, nor the probability that the measurements in each bin could fluctuate towards the expectation value of the broader radio population. Therefore, the appropriate confidence intervals are attached to the sky model, not to the individual data points (for which the observed counts, median and sum have zero or negligible uncertainties for the purpose of this analysis), and it is the degree to which the latter fall outside the former that provide the measure of statistical significance which is appropriate for our aims. 

We estimated the expectation values and confidence intervals for the broader population statistics differently depending on the statistic in question. For the polarised source counts (plotted in the uppermost panel in Figure \ref{fig:counts_plus_flux_in_annular_bins_mod3}), we assume the probability distribution corresponds to a spatial Poisson point process with an expected source count of 27 polarised sources per square-degree --- this value being the average polarised source density outside 2 degrees radius from the cluster centre. The confidence intervals then follow from the standard properties of Poisson processes. For the sums and medians of the polarised and total flux in each annular bin (middle and lower panels of Figure \ref{fig:counts_plus_flux_in_annular_bins_mod3}), we derived the confidence intervals from a bootstrap analysis of our full sample. That is, we randomly select (with replacement) $10^5$ independent sub-samples of 76 radio sources --- the number of polarised sources closer than 1 degree to the cluster centre --- situated further than 2 degrees from the cluster centre in the mosaic. The relevant confidence intervals follow from the percentiles of these data. Ideally, we should also impose the constraint that each sample be taken from a contiguous $\pi$ square degree area, but our mosaic is too small to provide a sufficient number of independent samples for this purpose. Moreover, we caution that our mosaic region may not be representative of the broader extragalactic sky, which would affect our calculated confidence intervals. Future ASKAP surveys are needed to provide crucial new information about the global clustering properties of radio sources, particularly in polarisation, at this combination of survey depth, resolution, and frequency (see Sections \ref{sec:discoreduced} and \ref{sec:concluding}). 



\bibliographystyle{pasa-mnras}
\bibliography{bibliography}

\end{document}